\documentclass[fleqn,usenatbib]{mnras}

\pdfoutput=1

\usepackage{newtxtext,newtxmath, tablefootnote}
\usepackage{mathptmx}
\usepackage{color}

\usepackage[T1]{fontenc}
\usepackage{ae,aecompl}
\usepackage{threeparttable}
\usepackage{caption}
\usepackage[utf8x]{inputenc}

\usepackage{graphicx}	
\usepackage{amsmath}	
\usepackage{amssymb}	
\usepackage{pdflscape}

\title[Mass content of UGC 6446 and UGC 7524]{Mass content of UGC 6446 and UGC 7524 through HI rotation curves: 
deriving the stellar discs from stellar population synthesis models}

\author[P. Repetto et al.]{
P. Repetto,$^{1}$\thanks{E-mail: prsatch6@gmail.com}
Eric E. Mart\'{\i}nez-Garc\'{\i}a$^{2}$,
M. Rosado$^{3}$
and R. Gabbasov$^{4}$
\\
$^{1}$Via Gallaretta 42, Castelletto d'Orba, Alessandria, Italia, C.P. 15060\\
$^{2}$Cerrada del Rey 40A, Chimalcoyoc Tlalpan, Ciudad de M\'{e}xico, C.P. 14630\\
$^{3}$Instituto de Astronom\'{\i}a, Universidad Nacional Aut\'{o}noma de M\'{e}xico, Circuito de la Investigaci\'on Cient\'{\i}fica, \\
Ciudad Universitaria, Ciudad de M\'{e}xico, C.P. 04510\\
$^{4}$Instituto de Ciencias B\'asicas e Ingenier\'{\i}as, Universidad Aut\'{o}noma del Estado de Hidalgo, Carretera Pachuca-Tulancingo, \\ 
Hidalgo, M\'{e}xico, C.P. 42184}

\date{Accepted XXX. Received YYY; in original form ZZZ}

\pubyear{2015}

\begin{document}
\label{firstpage}
\pagerange{\pageref{firstpage}--\pageref{lastpage}}
\maketitle

\begin{abstract}

In this work we study the mass distribution of two irregular 
galaxies, UGC 6446 and UGC 7524, by means of HI rotation curves
derived from high resolution HI velocity fields obtained through
the Westerbork Synthesis Radio Telescope data archive. We constrain
the stellar and gas content of both galaxies with stellar population 
synthesis models and by deriving the HI+He+metals rotation curves from
the total HI surface density maps, respectively. The discrepancy between
the circular velocity maxima of the stellar plus the HI+He+metals 
rotation curves and the observed HI rotation curves of 
both galaxies requires the inclusion of a substantial amount of dark matter.
We explore the Navarro Frenk and White, Burkert, Di Cintio, 
Einasto and Stadel dark matter halo models. We obtain acceptable
fits to the observed HI rotation curves of UGC 6446 and UGC 7524 with 
the cored Burkert, Einasto and Stadel dark matter halos. In particular,
Einasto and Stadel models prove to be an appropriate alternative to the 
Burkert dark matter halo. This result should increase the empirical basis
that justify the usage of dark matter exponential models to adjust the 
observed rotation curves of real galaxies.

\end{abstract}




\section{Introduction}\label{sec:sc0}

The rotation curves (RCs) of galaxies trace the circular velocities 
of the total galactic density distribution as a function of radius.
Assuming centrifugal equilibrium it is possible to gather information 
about the mass of the most important components rotating under the 
influence of the central gravitational potential. The principal gravitating constituents
in a galaxy are stars, gas, baryonic dark matter (BDM) (mainly
stellar remnants, substellar objects (with masses between 0.01 and 0.08 
solar masses \citep{Fukugita2004}), cold and warm molecular gas and cold neutral gas), and 
non-baryonic dark matter (DM). According to \citet{Read2014} the current amount of stars, gas
and BDM in the Galaxy within a radius of $\sim$ 8 kpc is M$_{\rm b}$ $\sim 1.1 \times 10^{10}$ 
M$_{\odot}$, considering the stellar mass M$_{*}$ $\sim  6.0 \times 10^9$ 
M$_{\odot}$ estimated by \citet{Bovy2012}, gas mass M$_{\rm HI+H2}$ $\sim 3.4 \times 10^9$ 
M$_{\odot}$ and BDM M$_{\rm BDM}$ $\sim 1.4 \times 10^9$ M$_{\odot}$ \citep{Flynn2006}. In external 
low redshift galaxies of morphological types from E to dE,
stars constitute $80\%$ of baryons, neutral and molecular gas form the remaining $20\%$. 
Other components such as halo stars, massive compact halo objects, hot ionized gas and dust 
represent a negligible fraction of the total baryons budget \citep{Read2005}. On the basis of 
these considerations, in the present work, we concentrate exclusively on the 
stellar and gaseous parts of the total baryonic content of galaxies.  

A common approach, to derive the stellar masses of observed galaxies, relies on
the stellar population synthesis (SPS) technique that reproduces a galaxy spectrum through the sum 
of the spectra of the stellar populations residing in the galaxy disc considering 
groups of stars of diverse age and distinct chemical content \citep{Walcher2011}. 
In the current literature one commonly used procedure to estimate 
the spectrum of a stellar population is the `isochrone synthesis` 
method introduced by \citet{Charlot1991} (CB91), whose principal ingredients are stellar evolutionary
tracks, libraries of stellar spectra, an initial mass function (IMF) and a prescription
for the star formation history. The CB91 strategy interpolates the 
stellar evolutionary tracks to generate stellar isochrones
and produce (from an IMF) the stellar mass distribution, for a given chemistry. 
The spectrum of a stellar population is obtained integrating over a certain 
mass range the spectra of the individual stars weighted for the determined stellar mass function. 
A galaxy spectrum is computed as a time integral of all stellar population spectra weighted for 
a particular star formation rate (SFR) recipe \citep{Courteau2014}. 

There are a number of uncertainties associated with each constituent of 
the SPS strategy that we concisely report here in order to properly address the limitations
and shortcomings of SPS methods. Isochrones are principally affected by errors connected to
the modelling of delicate phases of stellar evolution, namely convection, rotation, mass-loss, 
binary evolution and thermal pulses of stars in the asymptotic giant branch 
\citep{Charlot1996, Conroy2009, Conroy2013}. Stars from 1.1 to 7 M$_{\odot}$ have 
convective cores that develop a certain degree of overshooting due to fluid elements with non
zero velocity moving from the radiative to the convective regions. Core convective overshooting
produces bluer stellar colors of $\approx$ 0.1 mag for stars in the age interval 
$\sim$ 0.1-1 Gyr, a brighter main sequence turnover and an increase of $\sim$25$\%$ of the main 
sequence lifetime \citep{Conroy2010}. Stellar rotation augments of a 25$\%$ the permanence of 
stars in the main sequence strip \citep{Maeder2012}. SPS colors are bluer ($\sim$
0.1-1.0 mag for different age and wavelength intervals) and stars luminosity is higher 
due to stellar rotation according to the analysis of \citet{Vazquez2007}. Mass-loss drastically 
changes the luminosity, mass, metallicity and lifetime of massive stars (> 40 M$\odot$) 
\citep{Conroy2013}. Binary evolution produces bluer SPS colors and increases the age and metallicity
of stars \citep{Zhang2005}. However, \citet{Hernandez2013} pointed out that the 
inclusion of binary stars in SPS models only affects the spectra of early-type galaxies in the 
ultraviolet (UV). Thermally pulsing asymptotic giant branch (TP-AGB) stars are moderately
cold high luminosity stars with masses between 1.5 and 8 M$_{\odot}$ that undergo a thin shell 
instability during the helium burning phase \citep{Schwarzschild1968}. The modelling of these stars
is crucial for SPS application to correctly derive mass-to-light ratios (M/L) and
metallicities of intermediate age stars \citep{Walcher2011}. The most important drawbacks of 
theoretical libraries of 
stellar spectra are the following: incorrect input of atomic and molecular parameters, inadequate 
treatment of convection, non local thermodynamic equilibrium and non plane parallel geometries and,
last but not least, faulty and fragmentary atomic and molecular line lists \citep{Kurucz2011}. On
the other hand the problems of the empirical stellar libraries originate from atmospheric absorption, 
flux calibration, poor spectral resolution and limited wavelength range and the modest span of the 
parameter space. The uncertainties related to the IMF could 
affect the spectral energy distribution of SPS and the global normalization of the stellar M/L.
Estimates of SFRs through spectral energy distribution fitting are problematic due to
the age-dust-metallicity degeneracy, the adopted dust model and star formation history libraries 
\citep{Conroy2013}. \citet{Salim2007} showed that the SFRs derived 
from UV-optical spectral energy distributions have systematic uncertainties of $\lesssim$ 0.3 dex. 
Some of the issues connected to critical phases of stellar evolution such as core convective overshooting, 
mass-loss and TP-AGB have been addressed and included in the state of the art SPS 
models. In general the SPS method is considered a cautious approach to determine 
the stellar masses of galactic systems, regardless of the downsides reported above.

The massive gaseous part of a galaxy is primarily composed of neutral (HI) and molecular hydrogen (H$_2$).
The proportion of these two gas phases in a galaxy disc depends mostly on redshift, SFR 
and the total stellar mass as outlined by some theoretical and observational studies whose principal findings
we briefly summarize below. \citet[LG14]{Lagos2014} analysed the look-back time dependence 
of the HI and H$_2$ spatial densities using the semi-analytic, $\Lambda$CDM 
\citep{Blumenthal1984} scenario, galaxy formation and evolution
code GALFORM \citep{Cole2000}. The authors found that the HI volume density is nearly constant for z $\lesssim$ 2 
and diminishes at higher z. The H$_2$ spatial density correlates with the SFR volume density, 
augments up to z $\lesssim$ 3 and shows a gradual decrease for z > 3. In addition the GALFORM study determined
that HI is predominant in galaxies with \mbox{SFR < 1 M$_{\odot}$ yr$^{-1}$} and stellar masses 
\mbox{M$_{\star}$ < 10$^9$ M$_{\odot}$}, whereas H$_2$ is more abundant in galaxies with \mbox{SFR > 10 
M$_{\odot}$ yr$^{-1}$} and M$_{\star}$ > 10$^{10}$ M$_{\odot}$. The results of other semi-analytic models 
with similar physical recipes are consistent with the conclusions of LG14 \citep{Popping2014, Fu2012}. 
Contrasted with observations, the results of LG14 at \mbox{z $\sim$ 0} are congruent with estimates of the 
current overall HI and H$_2$ densities performed respectively by \citet{Martin2010} using data from the Arecibo
Legacy Fast ALFA survey (ALFALFA, \citealt{Giovanelli2005}) and by \citet{Keres2003} employing data
from the Five College Radio Astronomy Observatory Extragalactic CO Survey \citep{Young1995}. At z $\lesssim$ 3 
measures of the HI volume density of damped Lyman-$\alpha$ systems in the optical waveband detected a small 
variation of the HI mass density with increasing redshift in accordance with the LG14 conclusions 
\citep{Noterdaeme2012}. At z $\lesssim$ 2 estimations of the H$_2$ mass density of \citet{Berta2013} are 
concordant with the predictions of LG14. The hot and warm ionized hydrogen, helium, metals and dust constitute 
a negligible mass portion of the total baryonic fraction of a galaxy as indicated by the study of \citet{Read2005} 
and we omit the discussion about these components accordingly. 

In the next paragraphs we concisely recapitulate
the most important observational and theoretical evidences of the existence of the DM, both on cosmological and 
astrophysical scales. From a cosmological perspective, the $\Lambda$CDM theory predicts with very high accuracy the 
cosmological parameters originated by the cosmic microwave background anisotropies in temperature and density, the 
baryonic acoustic oscillation and the formation and evolution of large scale structures \citep{Planck2014}. Some 
concerns arise on galactic scale from the comparison between the simulated and the observed DM profile. 
The cored DM profile measured from observations contrasts with the theoretical prediction of a 
cuspy profile. On the theoretical ground some solutions have been proposed to lessen this discrepancy.
In the following we recapitulate solely the most relevant hypothesis. \citet{Governato2012} investigated 
a scenario at different redshifts, previously devised by \citet{Governato2010}, in which several gas 
outflows originated by winds from supernova convey energy to the DM particles smoothening the originally 
simulated cuspy DM profile into a cored one with an inner slope concordant with that observed in dwarf galaxies.
\citet{Nipoti2015} proposed a mechanism where a galactic disc (formed by the gas infall 
into the DM halo) disrupts into large compact masses that transfer their energy through dynamical friction to the DM particles and consequently softens the initial DM cusp into a core
with an inner slope close to zero. \citet{Maccio2015} examined a coupling between cold DM and dark energy with an additional separated 
warm dark matter component. The authors found the usual cuspy DM profile on galactic mass scales ($\sim$10$^{12}$ M$_{\odot}$), whereas 
on satellite mass scales ($\sim$10$^9$ M$_{\odot}$) their simulations develop DM cored profiles with radii of hundreds of parsecs as 
observed for the Galaxy satellites. From an observational perspective \citet{deBlok2010} reviewed, among other things, possible effects 
related to the observation process that could affect the determination of the DM density profile. In particular low resolution HI and 
H$\alpha$ long-slit observations suffer low spectral resolution, telescope pointing problems and difficult to quantify non-circular 
motions that could bias the actual slope of the DM density profile. The author concludes that high resolution H$\alpha$ and HI velocity 
fields, nowadays largely available, substantially decrease the consequences of the observational downsides of the earliest RCs observations 
leading to a correct determination of the DM profile slope. Regardless the copious theoretical and observational efforts in the last decades, 
the cuspy/core problem is outstanding and seems hard to address in its totality. 

In this paper we continue the study of the total mass distribution in irregular galaxies that we 
begin in \citet[RP15]{Repetto2015} (see also \citealt{Repetto2013}, RP13 for a more extensive presentation of the
adopted methodology), analysing the total mass distribution of the two irregular galaxies UGC 6446 and UGC 7524 
performing a stellar population synthesis study to obtain the stellar contribution to the total mass of these two 
galaxies and deriving the gaseous part of the total mass from the observed HI distribution.
We use optical, near infrared photometry and HI surface brightness to account for the stellar and gas
mass of UGC 6446 and UGC 7524, respectively. The observed HI RCs provide the total gravitating mass. The election of these 
two galaxies is dictated by the fact that the velocity fields present negligible deviations from circular motions, as 
determined by the analysis of the observed radial velocity maps reported in section~\ref{sec:sc3}. The corresponding 
observed HI RCs derived in the same section are consequently suitable to obtain the DM halo and total masses of
UGC 6446 and UGC 7524. We test five DM models in an attempt to determine which kind of DM velocity profile adequately fits 
the DM RCs of UGC 6446 and UGC 7524. Our strategy parametrizes the DM component alone because the stellar and gaseous velocity
profiles are derived directly from observations, independently from the fitting procedure of the DM velocity profile. 
The determination of the DM amount is then reduced to the subtraction of the baryonic circular velocity 
components form the observed RC. 

This work is structured in the following manner: in section~\ref{sec:sc1} we perform 2D isophotal analysis 
of the photometric discs of UGC 6446 and UGC 7524 to obtain the number and type of stellar components present in their
discs. In section~\ref{sec:sc2} we estimate the stellar and neutral gas plus helium plus metals RCs. In section~\ref{sec:sc3} we 
derive the observed HI RCs and DM mass content of the two irregular galaxies subject of this article.
In section~\ref{sec:sc4} we analyse the fits to various DM halo models. Section~\ref{sec:sc5} is dedicated to discuss 
possible issues related to our DM estimates. The conclusions are highlighted in section~\ref{sec:sc6}

\section{UGC 6446 and UGC 7524 Galfit 2D isophotal analysis}\label{sec:sc1}

UGC 6446 is morphologically classified in the EFIGI catalogue \citep{Baillard2011} as a transition between spiral 
and irregular with very weak bulge and no obvious sign of an ordered spiral structure. The NED and 
HYPERLEDA\footnote{http://leda.univ-lyon1.fr/} \citep{Makarov2014} morphological types of UGC 6446 are respectively
Sd and Scd concordant with the EFIGI catalogue. \citet[MCD09 hereafter]{McDonald2009} performed 1D bulge-disc 
decomposition of UGC 6446 , among other 65 spiral galaxies in the Ursa Major cluster of galaxies, using K$^{\prime}$
band photometry from \citet{Tully1996}, finding that UGC 6446 has a very faint bulge fitted by a Sersic
profile \citep{Sersic1968}. We accomplished the 2D bulge-disc decomposition of UGC 6446 with the GALFIT software 
\citep{Peng2002, Peng2010} considering a Sersic profile for both bulge and disc. We performed the 2D Sersic disc 
fit on the Hubble Space Telescope (HST) F606W image, and the 2D Sersic bulge fit on the 2MASS 
J band image \citep{Skrutskie2006}. The resultant parameters are listed in Table~\ref{tab:tb1}.

\begin{table}
\centering
\begin{threeparttable}
\caption{GALFIT results for UGC 6446 and UGC 7524.}
\label{tab:tb1}
\begin{tabular}{p{1.5cm}p{0.5cm}p{0.5cm}
p{0.5cm}p{0.5cm}p{0.8cm}p{0.8cm}}
\hline
UGC 6446 & mag\tnote{(1)} & r$_{\rm s}$\tnote{(2)} & 
n$_{\rm s}$\tnote{(3)} & q\tnote{(4)} & pa\tnote{(5)}\\
\hline
Disc & 18.81 & 1.70 & 0.8 & 0.76 & 22.44\\
Bulge & 17.22 & 0.58 & 0.44 & 0.53 & 25.77\\
\hline
UGC 7524 & mag & r$_{\rm s}$ & n$_{\rm s}$ & q & pa\\
\hline
Disc & 19.76 & 3.93 & 2.56  & 0.61  & -42.71\\
\hline
     & $\mu_{\rm 0}$\tnote{(6)} & r$_{\rm out}$\tnote{(7)} & $\alpha$\tnote{(8)} & q & pa\\
\hline
Bar & 25.3 &  3.1 & 2.26 & 0.12 & -59.75\\
\hline
\end{tabular}
\begin{tablenotes}
\item[1] Apparent magnitude
\item[2] Disc scale length (kpc)
\item[3] Sersic index
\item[4] Axis ratio
\item[5] Position angle (degrees)
\item[6] Central Surface Brightness (mag arcsec$^{-2}$)
\item[7] Outer truncation radius (kpc)
\item[8] Outer truncation sharpness (Central slope $\beta$=0)
\end{tablenotes}
\end{threeparttable}
\end{table}

The morphological classification of UGC 7524 (NGC 4395) is SA(s)m in the NED database, SABdm 
\citep{Ann2015} in the HYPERLEDA and SIMBAD \citep{Wenger2000} database. We accomplished the 
GALFIT 2D disc isophotal analysis of UGC 7524 employing a Sersic profile for the disc. The Sersic function was 
adjusted to the SDSS DR7 \citep{Frieman2008} g band image. We fit
the bar component with a Ferrers profile \citep{Ferrers1877} adjusted to the 3.6$\,\mu$m IRAC image 
of the Spitzer infrared satellite. The results of the 2D disc and bar fits 
are tabulated in Table~\ref{tab:tb1}. The stellar components detected in this
section for UGC 6446 and UGC 7524 are included in the stellar mass models presented in the next sections.

\section{Baryonic RC\lowercase{s} estimates}\label{sec:sc2}

\subsection{Stellar RCs}

The use of stellar population synthesis (SPS) models~\citep[e.g.,][BC03]{bru03},
as a mean to convert stellar light 
into mass (through M/L) has been
frequently promoted~\citep[e.g.,][]{bel01,bel03}.
Notwithstanding its common degeneracies (e.g. age-metallicity-dust)
SPS models can assess reliable mass estimates in general.
Different studies have acknowledged that the main systematic uncertainty
in mass estimates via SPS fitting comes from different treatments of the TP-AGB phase~\citep{mara06,bru07,kan07,Conroy2009}.
For instance, the BC03 models adopt the physical properties of TP-AGB stars from~\citet{vas93},
while the 2007 version of Bruzual \& Charlot SPS models (CB07) uses the~\citet{mari07} prescription.
The CB07 models produce brighter near infrared (NIR) magnitudes, in consequence
the CB07 mass estimates are lower, by a factor of $\sim2$, when compared to BC03.
 
One innovative technique in mass estimates is the ``resolved stellar mass map method''~\citep[][ZCR]{zcr09}, 
which delivers a map of the stellar mass surface density by photometric means. 
Galaxy masses determined by unresolved means (where galaxies are treated as point sources) are subject to 
underestimations, thus the need for resolved structures~\citep{zcr09,sor15}.
Despite its potential, the ZCR method produces biased spatial structures which contradict the
expectations from NIR imaging. The bias consists of a morphology with filaments, 
and a spatial coincidence between dust lanes and stellar mass surface density. 
The bias is due to a limited M/L accuracy ($\sim0.1-0.15$ dex) arising 
from uncertainties inherent to observations, and degeneracies in SPS models 
\citep[]{mar16a}.

\subsubsection{The Bayesian successive priors (BSP) algorithm}

The BSP algorithm \citep[]{mar16a}
is intended to solve for the bias in the ZCR method.
The innovation consists in the use of prior information regarding the spatial distribution
of the stellar surface mass density as deduced from observations.
The massive older population of a galaxy is mainly traced in the NIR bands,
specially the $K$-band~\citep[][]{rix93}.

The BSP algorithm is intended to work with surface photometry, and a Monte Carlo SPS library
with several templates. Taking into account adequate probability distributions
for the star formation history (SFH), the stellar metallicity, and the dust content,
the SPS library is built using Montecarlo methods~\citep[e.g.,][]{daC08,mag15}.
The BSP algorithm is independent of the choice of the SPS library. 
The algorithm consists of three iterations, and is applied on a pixel-by-pixel basis
by using Bayes' theorem

\begin{equation}~\label{Bayes}
  P\left({\rm \frac{M}{L}} \mid C\right) = \frac{P\left(C \mid {\rm \frac{M}{L}}\right)~P\left({\rm \frac{M}{L}}\right)}{P\left(C\right)},
\end{equation}

\noindent
where $P\left({\rm \frac{M}{L}} \mid C\right)$ is the probability of having M/L
for a certain stellar population if colours $C$ are observed from the photometry,
i.e., the {\it{posterior}} M/L probability. 
All the M/L values in the BSP algorithm correspond to a NIR band, e.g., the $K$-band.
The {\it{prior}} M/L probability distribution is given by $P\left({\rm \frac{M}{L}}\right)$, and represents
any previous knowledge we may have about the expected M/L.
$P\left(C\right)$ is a normalisation constant.
The probability $P\left(C \mid {\rm \frac{M}{L}}\right)$ is usually determined by fitting the observed 
colours to the colours of the SPS library templates via $\chi^2$ minimisation,
or maximum likelihood methods.
In these cases the prior probability distribution, $P\left({\rm \frac{M}{L}}\right)$, is assumed
to be flat, i.e., each template in the SPS library has the same probability
to be fitted. This does not considers the probability distributions of the SFH, metallicity,
or dust content inherent to the construction of the SPS library.
The novelty in the BSP algorithm consists of using a non-flat prior probability distribution, $P\left({\rm \frac{M}{L}}\right)$,
when fitting the observed colours to the SPS library templates.
The three iterations in the BSP algorithm are as follows.

\begin{enumerate}

\item In the first iteration we use a flat prior probability distribution, i.e., $P\left({\rm \frac{M}{L}}\right)=$ constant.
Then we fit the observed colours to the templates in the SPS library
via a maximum likelihood estimate, i.e.,
maximising the probability

\begin{equation}
P\left({\rm \frac{M}{L}} \mid C\right) \propto\exp \left(-\frac{\chi^{2}}{2}\right).
\end{equation}

Until this point the algorithm is similar to the ZCR method.
Then we identify all the pixels where the difference between the observed colours,
and the SPS library colours, is less than $3\sigma_{\rm phot}$, where $\sigma_{\rm phot}$
is the uncertainty in the observed photometry.
This guarantees that each pixel can be described by at least one template in our SPS library.
Next we use the posterior NIR M/L from all the fitted pixels and calculate the statistical median.

\item In the second iteration the median NIR M/L from iteration number 1
is used as a constant parameter, i.e., ${\rm \left(\frac{M}{L}\right)}^{\rm prior}=\rm{constant}$, throughout the disc.
The $P\left({\rm \frac{M}{L}}\right)$ probability distribution 
function (PDF) is now assumed to be a Gaussian of the form

\begin{equation}~\label{prior}
P\left({\rm \frac{M}{L}}\right) \propto \exp \left(-\left[\frac{{\rm \left(\frac{M}{L}\right)}^{\rm prior} - {\rm \left(\frac{M}{L}\right)}} 
{\sigma_{\rm \left(\frac{M}{L}\right)}} \right]^{2}\right),
\end{equation}

\noindent where M/L is determined from the fitted templates, and 
${\sigma_{\rm \left(\frac{M}{L}\right)}}$ is the photometric uncertainty in M/L.
Under these new assumptions we fit all the pixels by using

\begin{equation}~\label{maxlike_and_prior}
P\left({\rm \frac{M}{L}} \mid C\right) \propto \exp\left(-\frac{\chi^{2}}{2}\right) P\left({\rm \frac{M}{L}}\right).
\end{equation}

Subsequently, we identify the pixels where the difference between the observed colours, and the SPS
library colours is less than $\sigma_{\rm cd}$. In this case, $\sigma_{\rm cd}$ is estimated
from the resulting ``colour differences'' between the observed colours of the pixels, and the colours of the fitted templates.
We compute the 16th and 84th percentiles\footnote{The number where a certain percentage of values fall below that percentile.}
belonging to the distribution of values obtained from these ``colour differences'', arising from all the
adopted pixels. Then we estimate $\sigma_{\rm cd}$ as the difference between the 84th percentile and the 16th percentile, divided by 2.
The identified pixels whose ``colour differences'' fulfill the $<\sigma_{\rm cd}$ condition will be the ``backbone''
of our mass map. The rest of the pixels belong mainly to relatively young luminous red stars,
low surface brightness regions in the outskirts of the disc, and high extinction regions.
For these pixels we used the information from the ``backbone'' pixels to interpolate the stellar
mass surface density, and obtain new NIR M/L ratios.
The interpolation is done with the purpose that the resolved mass map resembles the NIR
spatial distribution of the old stellar disc.

\item The third and last iteration is intended to deal only with the interpolated pixels in
iteration number 2. We apply Bayes' theorem with the new NIR mass-to-light ratios which are used as 
independent parameters, ${\rm \left(\frac{M}{L}\right)}^{\rm prior}$, in equation~\ref{prior}.
We adopt a pixel-by-pixel treatment, and after fitting via equation~\ref{maxlike_and_prior},
we obtain a bias-free stellar mass surface density map.

\end{enumerate}

\begin{figure*}
\centering
\includegraphics[width=1.0\hsize]{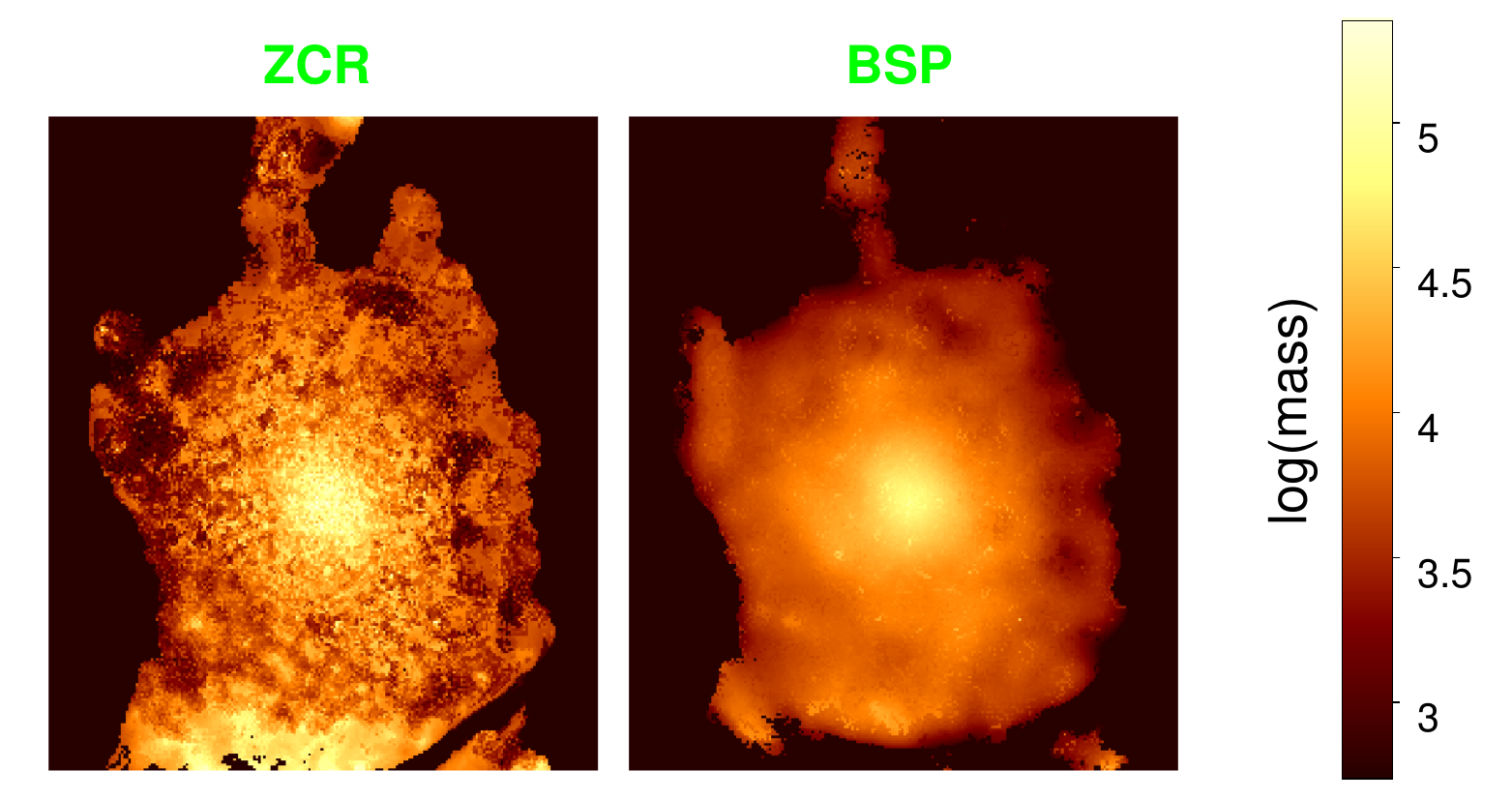}
\caption[f1.ps]{
Resolved map of the stellar mass surface density for UGC~6446.
{\it{Left}}: with the method of~\citet[][ZCR]{zcr09}.
{\it{Right}}: with the Bayesian successive priors (BSP) algorithm \citep[]{mar16a}.
Stellar mass is given in $M_{\sun}$~pixel$^{-1}$.
~\label{fig1}}
\end{figure*}

\begin{figure*}
\centering
\includegraphics[width=1.0\hsize]{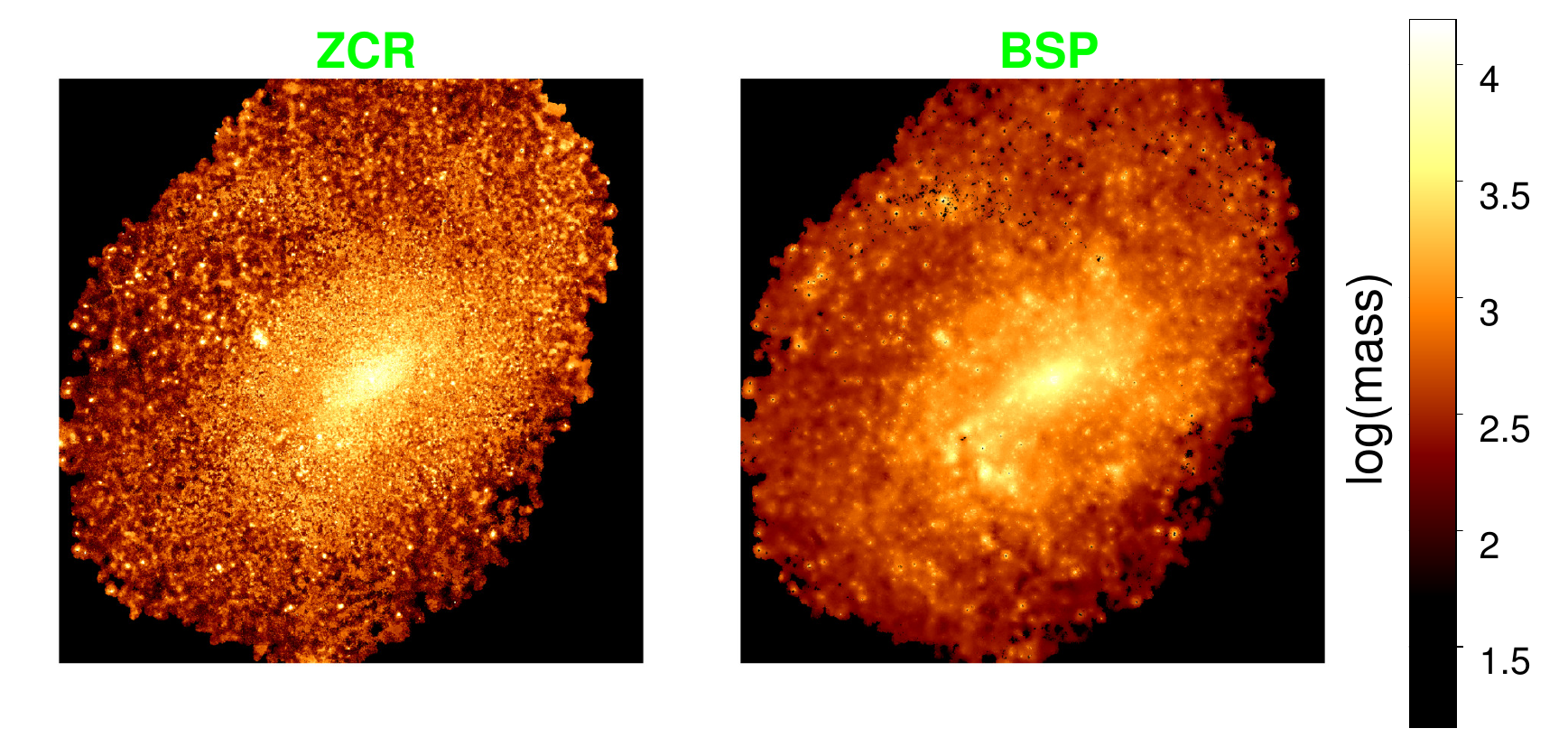}
\caption[f2.ps]{
Same as Figure~\ref{fig1} for UGC~7524.
~\label{fig2}}
\end{figure*}

\begin{figure*}
\centering
\includegraphics[width=1.0\hsize]{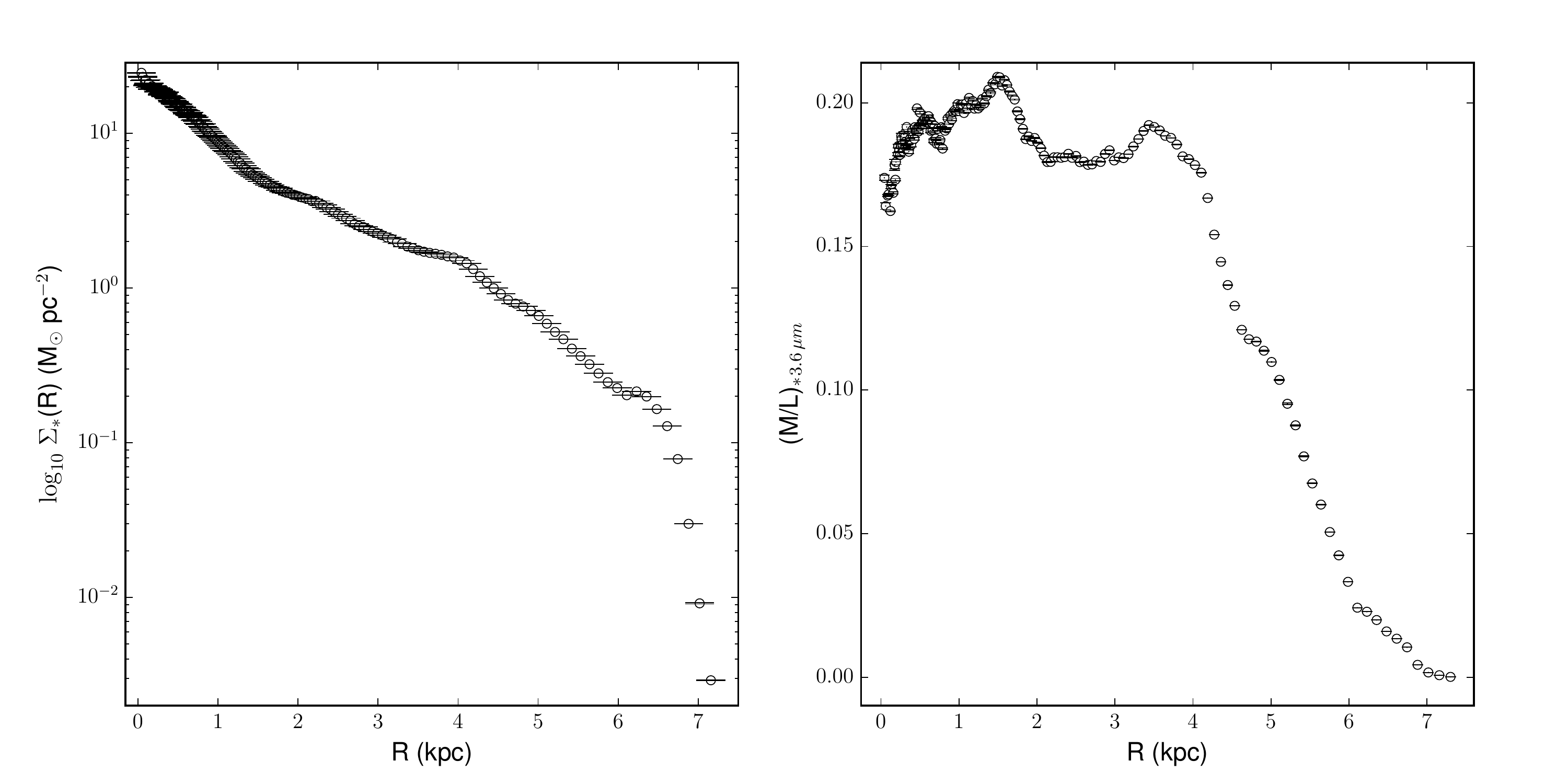}
\caption[f3.pdf]{
Stellar radial profiles of UGC~6446.
{\it{Left}}: Azimuthal average of the stellar mass surface density, $\Sigma$, as a function of radius.
{\it{Right}}: Azimuthal average of the stellar mass-to-light ratio, ${\rm \left(\frac{M}{L}\right)}_{3.6\,\micron}$, as a function of radius.
~\label{fig3}}
\end{figure*}

\begin{figure*}
\centering
\includegraphics[width=1.0\hsize]{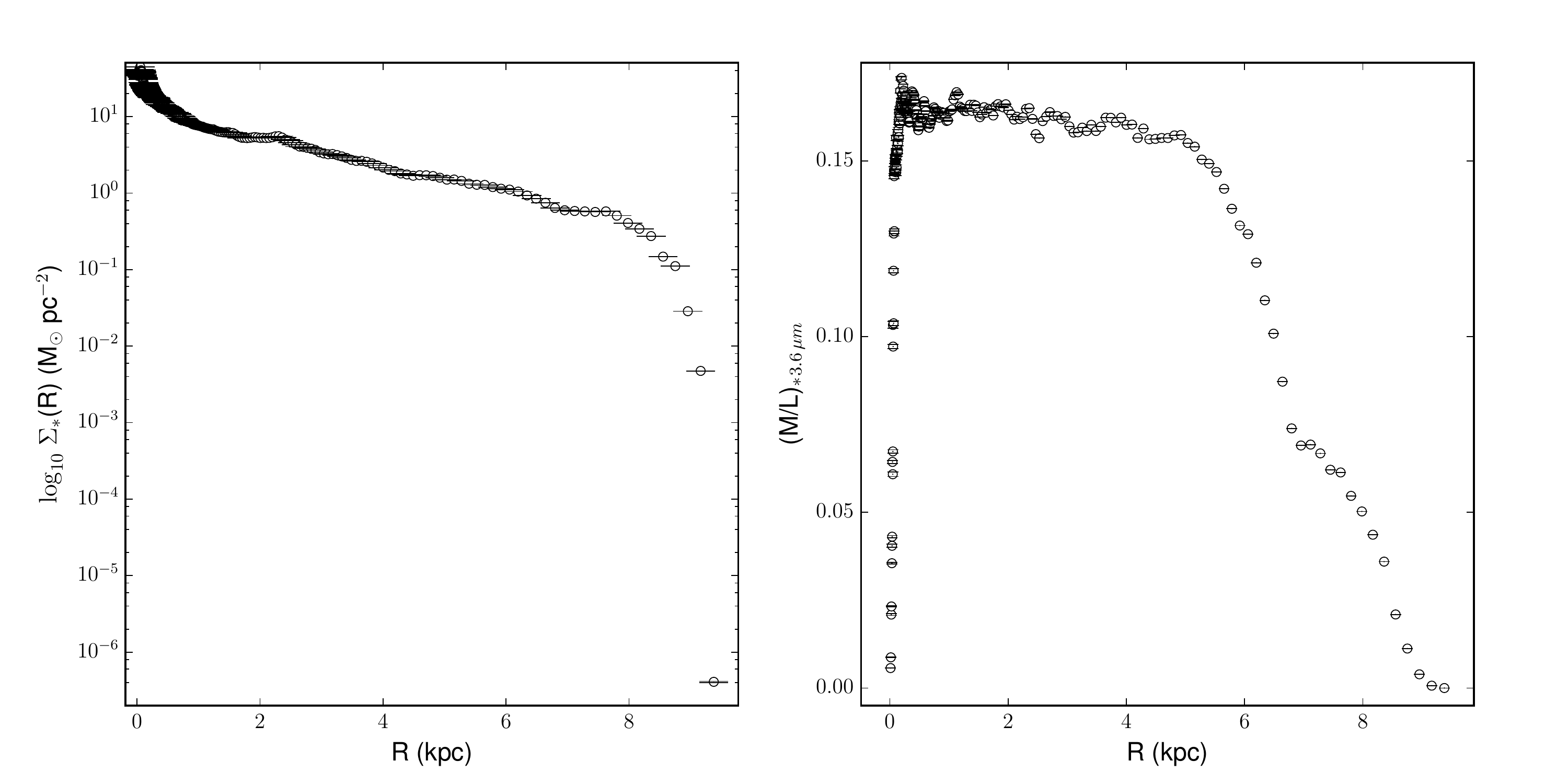}
\caption[f4.pdf]{
Same as Figure~\ref{fig3} for UGC~7524.
~\label{fig4}}
\end{figure*}

In this paper we apply the BSP algorithm to UGC 6446 and UGC 7524. We used SDSS DR12~\citep{ala15},
$g$, and $i$-band photometry, as well as $3.6\,\micron$ IRAC channel 1 imaging from
``The Spitzer Survey of Stellar Structure in Galaxies''~\citep[S$^4$G,][]{she10}.
SDSS data were registered and resampled to the S$^4$G data. The {\tt{Adaptsmooth}}~\citep{zbt09} code
was used to increase the S/N ratio in the outskirts of the discs.
We adopt the optical-NIR branch of the MAGPHYS\footnote{http://www.iap.fr/magphys/magphys/MAGPHYS.html} library~\citep{daC08},
which includes the CB07 SPS models.
The resulting resolved mass maps are shown in Figures~\ref{fig1} and~\ref{fig2}, 
for UGC 6446 and UGC 7524 respectively.
In these figures we also compare the results obtained by applying the
ZCR method where we notice a bias in the spatial structure. This bias is even more
evident for normal spirals with prominent dust lanes~\citep[see, e.g.,][]{mar16a}.
For the objects analysed, we found that by applying ZCR, instead of the BSP algorithm,
the integrated stellar masses are overestimated by a factor of $\sim1.5$.
In Figures~\ref{fig3} and~\ref{fig4} we show the azimuthal averages of the stellar mass surface density,
$\Sigma$, and the stellar mass-to-light ratio, ${\rm \left(\frac{M}{L}\right)}_{3.6\,\micron}$, as a function of radius, for
UGC~6446 and UGC~7524 respectively. We assume a distance of \mbox{12$\pm$1 Mpc} and 3.5$\pm$0.4 Mpc for UGC~6446
and UGC~7524, respectively. The deprojection parameters are those presented in Table 2 (present work). 

At this point, it is important to stress that in order to produce the stellar mass surface
densities curves displayed in the left panels of Figure~\ref{fig3} and~\ref{fig4} we used the whole M/L curves
showed in the right panels of the same figures. In addition we compute the total stellar mass and luminosity
at 3.6$\,\mu$m of UGC 6446 and UGC 7524 from the corresponding 2D maps adding the contribution of each single pixel.
We obtain a total stellar mass of 2.22$\times$10$^8$ M$_{\odot}$ and a total luminosity at 3.6$\,\mu$m of 
1.14$\times$10$^9$ L$_{\odot}$ for UGC 6446. For UGC 7524 the total stellar mass is 3.82$\times$10$^8$ M$_{\odot}$
and the total stellar luminosity at 3.6$\,\mu$m is 2.41$\times$10$^9$ L$_{\odot}$. From these estimations we derive 
an average ${\rm \left(\frac{M}{L}\right)}_{3.6\,\micron}$ of 0.19 and 0.16 for UGC 6446 and UGC 7524 respectively. 

The stellar RCs of UGC 6446 and UGC 7524 (see Figure~\ref{fig6}) are obtained from the stellar 
mass surface densities of Figures~\ref{fig3} and~\ref{fig4} following the recipe of \citet{Casertano1983} 
as implemented in the ROTMOD procedure of the Groningen Imaging  Processing System 
(GIPSY\footnote{http://www.astro.rug.nl/$\sim$gipsy/})~\citep{Allen1985, vanderHulst1992}, that 
calculates the RC for a certain density law.

\subsection{HI+He+metals RCs}\label{sec:sc22}

We obtain the HI full resolution ($\sim 12^{\prime\prime}$ $\times 12^{\prime\prime}$) total density maps 
(as well as velocity fields) of UGC 6446 and UGC 7524 from the 
WHISP\footnote{http://wow.astron.nl} (Westerbork HI Study of Spirals) survey 
database \citep{Swaters2002}. We derive the neutral gas surface density of UGC 6446 and UGC 7524 from the HI 
total density maps of both galaxies by means of the ELLINT task of the GIPSY package, using the deprojection parameters 
of \citet[see Table~\ref{tab:tb2}]{vanEymeren2011a}. 

\begin{figure*}
\centering
\includegraphics[width=1.0\hsize]{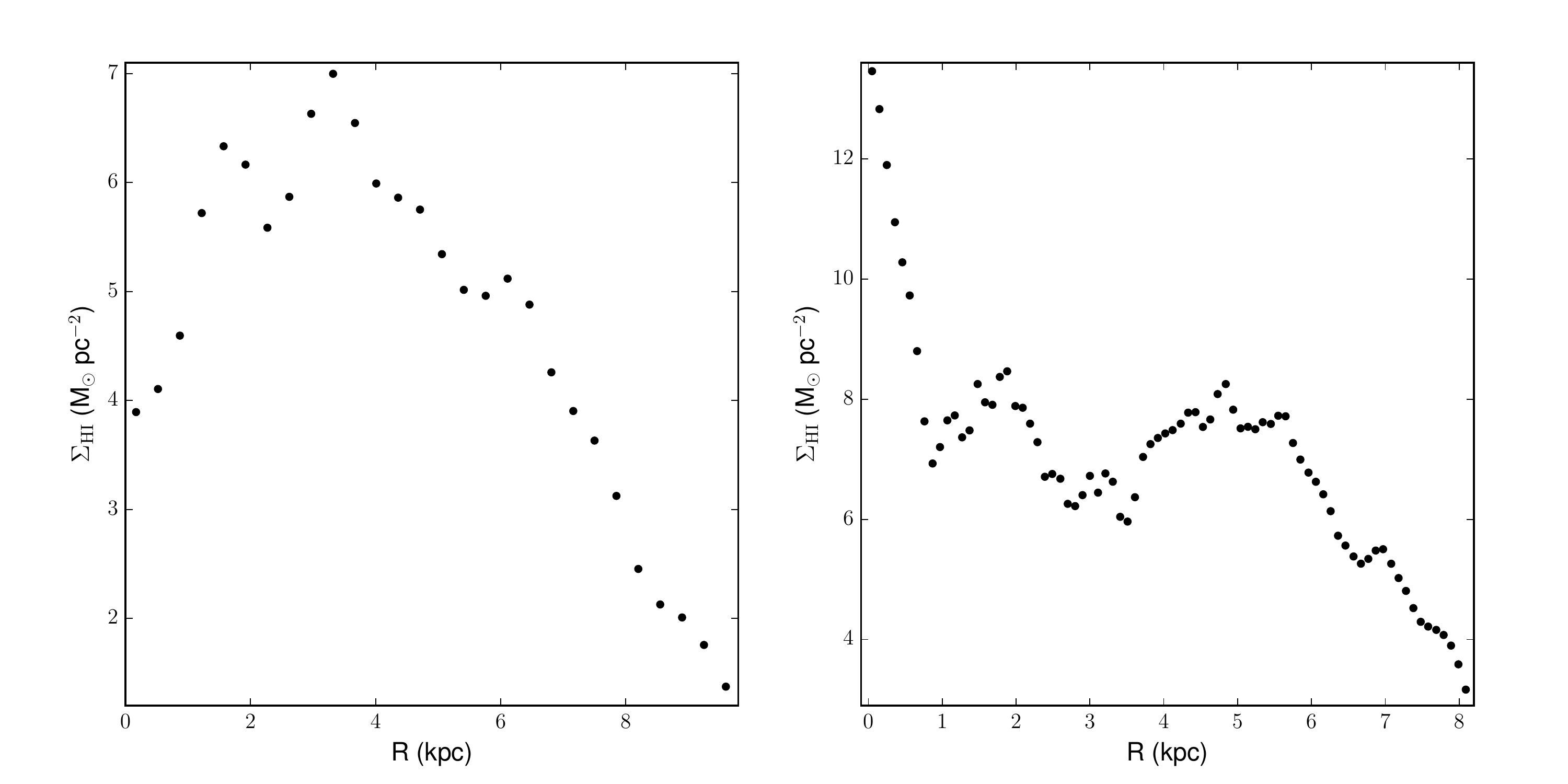}
\caption[f5.pdf]{
HI gas surface densities profiles of UGC~6446 and UGC~7524.
{\it{Left}}: Azimuthal average of the HI mass surface density of UGC~6446, $\Sigma_{\rm HI}$, as a function
of radius.
{\it{Right}}: Azimuthal average of the HI mass surface density of UGC~7524, $\Sigma_{\rm HI}$, as a function 
of radius.
Deprojection parameters as in Table~\ref{tab:tb2} (present work).
~\label{fig5}}
\end{figure*}

The ELLINT procedure computes the HI surface density profile dividing the HI total density map into elliptical concentric rings,
computing the mean surface density in each ring, and summing the latter quantity over the total number of annuli. The mean surface 
density is the azimuthal average of the flux in each pixel of a given ring under the supposition of a linear relation between flux
and mass. The gas circular velocities are built from the surface density profiles of both galaxies with the ROTMOD routine. The 
HI+He+metals RCs of UGC 6446 and UGC 7524 are atomic hydrogen RCs multiplied by a factor $1.33$. The HI surface density profiles of
UGC 6446 and UGC 7524 are displayed in Figures~\ref{fig5}. The HI+He+metals RCs for both galaxies are shown in Figure~\ref{fig6}.

\section{Observed HI and DM RC\lowercase{s}}\label{sec:sc3}

We employ a tilted rings fitting method through the GIPSY routine ROTCUR to generate the observed HI RCs 
of UGC 6446 and UGC 7524 from the full resolution velocity fields of both galaxies, using the kinematic parameters of 
\citet{vanEymeren2011a} as first guess values for the ROTCUR fit. 

We assume pure circular motion for both galaxies and fit the approaching and receding sides of both velocity fields 
keeping fixed the rotation velocities and allowing to vary in turn the position angle and inclination. We note no significant 
radial variation of the position angles and inclinations of both galaxies therefore we fixed those parameters close to their initial 
values. Next we let vary the rotation velocities of UGC 6446 and UGC 7524 determining the approaching and receding RCs of both 
galaxies. The residual velocity fields determined through the subtraction of the synthetic circular velocity fields from the observed velocity
fields of both galaxies for the approaching and receding sides show velocities smaller than 15 km s$^{-1}$ and the velocity difference between 
the approaching and receding residual velocity field of UGC 6446 is less than 1 km s$^{-1}$ whereas for UGC 7524 this discrepancy is 
$\sim$ 2 km s$^{-1}$. 
\begin{table}
\centering
\begin{threeparttable}
\caption{Kinematic and lopsidedness parameters of UGC 6446 and UGC 7524}
\label{tab:tb2}
\begin{tabular}{p{2.2cm}p{2.2cm}p{2.2cm}}
\hline
\hline
Present work &  UGC 6446 & UGC 7524\\
\hline
\hline
$\alpha$ (hms)\tnote{(5)} & 11 26 38.2 & 12 25 51.5\\
\hline
$\delta$ (dms)\tnote{(5)} & 53 44 51.5 & 33 32 13.7\\
\hline
V$_{sys}$\tnote{(6)} & 646.4 & 318.3\\
\hline
P.A.\tnote{(7)} & 191.3 & 325.1\\
\hline
INCL.\tnote{(8)}& 50.6 & 47.2\\
\hline
$\epsilon_{kin}$\tnote{(9)} & 0.004 & 0.003\\
\hline
\hline
VE\tnote{(10)} &  UGC 6446 & UGC 7524\\
\hline
\hline
$\alpha$ (hms)\tnote{(5)} & 11 26 40.7 & 12 25 50.4\\
\hline
$\delta$ (dms)\tnote{(5)} & 53 44 51.0 & 33 32 35.0\\
\hline
V$_{sys}$\tnote{(6)} & 646.0 & 317.0\\
\hline
P.A.\tnote{(7)} & 191.1 & 324.9\\
\hline
INCL.\tnote{(8)}& 50.5 & 46.9\\
\hline
$\epsilon_{kin}$\tnote{(9)} & 0.006 & 0.002\\
\hline
\hline
\end{tabular}
\begin{tablenotes}
\item[(5)] Kinematic centre (J2000).
\item[(6)] Systemic velocity (km s$^{-1}$).
\item[(7)] Position Angle (degrees).
\item[(8)] Inclination (degrees).
\item[(9)] Kinematic lopsidedness strength \citep{Jog2002}.
\item[(10)] \citet{vanEymeren2011a}.
\end{tablenotes}
\end{threeparttable}
\end{table}

The derivation of the RCs for the entire velocity fields of both galaxies proceeds with the determination of the kinematic centres and
systemic velocities respectively, fixing the remaining parameters. The final step consists in the derivation of the RCs for both halves 
of the velocity field of \mbox{UGC 6446} and UGC 7524. The corresponding RCs best fit parameters and the comparison with \citet{vanEymeren2011a} are reported in Table~\ref{tab:tb2}. The errors associated to the inclinations, position angles
and systemic velocities derived in the present work, are less than 1$^{\circ}$ and 1 km s$^{-1}$ respectively.

We estimate the intensity of the kinematic lopsidedness of UGC 6446 and UGC 7524 utilizing the relation of \citet{Jog2002}, that considers
the kinematic lopsidedness as a perturbation induced by a lopsided DM halo gravitational potential on an initial axisymmetric galactic disc. 
The determined lopsidedness parameters for \mbox{UGC 6446} and UGC 7524 are in good agreement with those obtained by \citet{vanEymeren2011a}. From our estimates we conclude that the degree of kinematic lopsidedness of UGC 6446 and UGC 7524 is small, consequently the velocity fields and the corresponding observed HI RCs are not so much affected by the lopsided perturbation. \citet{Swaters2002} classified UGC 6446 and UGC 7524 as objects with a strong kinematic lopsidedness. This classification 
is based on a qualitative ``by eye inspection'' of the observed velocity fields of both galaxies. We use a more quantitative criterion to determine the degree of kinematic lopsidedness of UGC 6446 and UGC 7524 and we reach opposite conclusions. The comparison with\citet{Swaters2002} is complicated by the fact that they do not provide quantitative information about the degree of the detected kinematic lopsidedness, whereas with the criterion of  \citet{Jog2002}, that we apply to UGC 6446 and UGC 7524, $\epsilon_{kin} \sim$0 reveals that the analysed galactic discs have very weak or absent kinematic lopsidedness.

\begin{figure*}
\centering
\includegraphics[width=1.0\hsize]{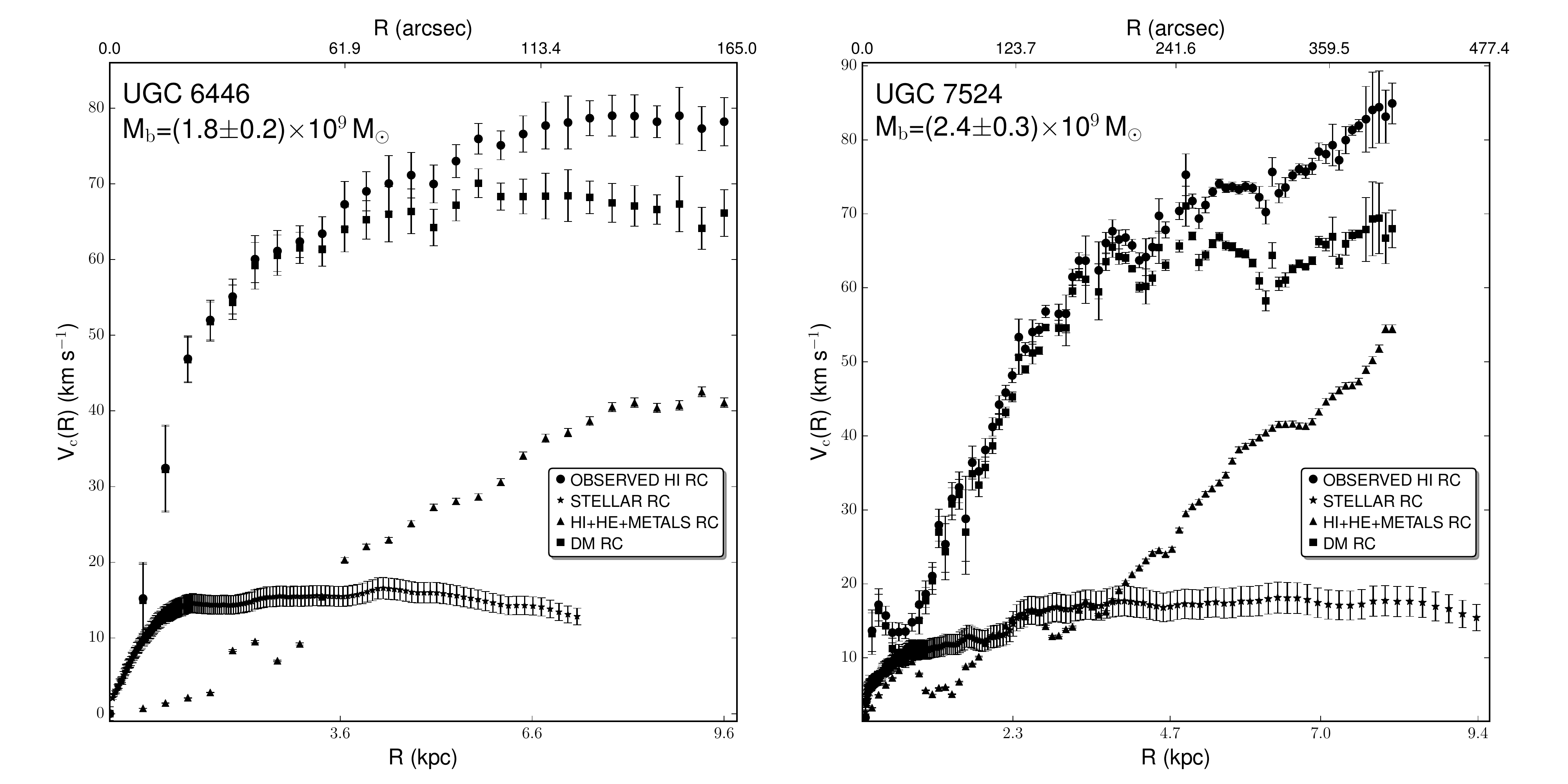}
\caption[f6.pdf]{
Observed HI RCs of UGC 6446 and UGC 7524 together with 
the stellar RCs, HI+He+metals RCs and DM RCs. The DM 
RCs and the associated errorbars are generated as outlined
in section~\ref{sec:sc3}. We use the deduced DM RCs to
perform the subsequent analysis.
~\label{fig6}}
\end{figure*}

The derivation of the stellar and gas RCs is described in section~\ref{sec:sc2}. The DM RCs of UGC 6446 and UGC 7524 and the associated error bars are calculated by means of the relation V$^2_{\rm h}$=V$_{\rm tot}^2$-(V$_{\star}^2$+V$_{\rm gas}^2$), where V$_{\rm h}$ are the DM halo RCs, V$_{\rm tot}$ are the observed HI RCs, V$_{\star}$ and V$_{\rm gas}$ represent the stellar and HI+He+metals RCs respectively. The error bars of the stellar and gas RCs are estimated considering the uncertainties related to each components separately, being $\pm$17$\%$ and $\pm$23$\%$ for the stellar RCs of UGC 6446 and UGC 7524, respectively. For the atomic gas RC of UGC 6446 we estimate an uncertainty of $\pm$3$\%$, as established through the propagation of the errors associated to the measures of the global neutral hydrogen flux by \citet{Verheijen2001}. For the neutral gas RC of UGC 7524 we estimate $\pm$2$\%$ as derived by means of the propagation of the errors relative to the overall neutral hydrogen flux of this galaxy estimated by \citet{Wolfinger2013}. Figure~\ref{fig6} displays the observed HI RCs, the baryonic RCs, and the DM RCs of UGC 6446 and UGC 7524. 

We estimate the total masses of the stellar discs of UGC 6446 and UGC 7524 through the relation M(r)=2$\pi$\,$\displaystyle\int_{0}^{r}$\,$\Sigma_{\star}$(R)\,R\,dR, where $\Sigma_{\star}$(R) represent the stellar surface densities of both galaxies displayed in the left part of figures~\ref{fig3} and~\ref{fig4}, and the upper limit of integration corresponds to 7.3 kpc and 9.4 kpc for UGC 6446 and UGC 7524 respectively. We employ the baryonic Tully-Fisher relation of \citet{McGaugh2012} to derive the total baryonic, gas, and neutral gas masses of UGC 6446 and UGC 7524. The adopted maxima of circular velocity are 79 km s$^{-1}$ and 85 km s$^{-1}$ for UGC 6446 and UGC 7524, respectively. The neutral hydrogen mass is provided by the expression \mbox{M$_{\rm HI}$=(M$_{\rm g}$/1.33)-M$_{\rm H2}$} of \citet{McGaugh2012} to correct the total gas mass M$_{\rm g}$ for the presence of helium and molecular hydrogen. Molecular hydrogen masses (see also section~\ref{sec:sc5}) were taken from \citet{Leroy2005}, \citet{Watson2012} and \citet{Boker2011}.

We obtain for UGC 6446 a stellar, total gas and neutral gas masses of (2.3$\pm$0.3)$\times$10$^8$ M$_{\odot}$, 
(1.6$\pm$0.2)$\times$10$^9$ M$_{\odot}$ and (1.2$\pm$0.2)$\times$10$^9$ M$_{\odot}$ respectively, and a baryonic total mass of 
(1.8$\pm$0.2)$\times$10$^9$ M$_{\odot}$. The stellar, total gas and neutral gas masses of 
UGC 7524 are (3.9$\pm$0.9)$\times$10$^8$ M$_{\odot}$, (2.1$\pm$0.3)$\times$10$^9$ M$_{\odot}$ and (1.3$\pm$0.3)$\times$10$^9$ M$_{\odot}$ 
respectively. The baryonic total mass of UGC 7524 is (2.4$\pm$0.3)$\times$10$^9$ M$_{\odot}$ It is important to note that 
the stellar masses of UGC 6446 and UGC 7524 derived in this section are completely congruent with the stellar masses estimated at the end 
of section~\ref{sec:sc2}.

\section{Fits to DM halo models}\label{sec:sc4}

We fit the DM RCs of the two galaxies under study to obtain an estimation of the total mass of UGC 6446 and UGC 7524 which is the sum of 
the DM RCs fit results and the stellar and neutral gas plus helium plus metals RCs. The resulting DM halos masses and the
corresponding total masses of both galaxies are presented in Table~\ref{tab:tb3}.
It is important to emphasise that the fits to DM RCs are models with two free parameters and represent the unique parametrization applied in this work because the stellar and neutral gas RCs are processed data gathered through observations. We describe below the fitting procedure adopted in the current analysis and outline the principal findings on the mass content of UGC 6446 and UGC 7524. Following RP15 we accomplish the DM mass derivation of UGC 6446 and UGC 7524 by means of the {\bf Minuit} curve fitting program within the ROOT\footnote{https://root.cern.ch} software package \citep{ReneBrun}. The Minuit routine provides two important tools to corroborate the validity of the DM RCs fitting solutions: an utility to scan the parameter space and a procedure to build the $\chi^2$ n-sigma contours of the final results. The $\chi^2$ n-sigma contours represent the ellipses of points describing the $n^2$ uncertainties of the fitted DM halo masses and radii and are an important instrument to evaluate the reliability of the solutions. The best results are expected to stay inside the innermost ellipse close to the centre. Another important quantity to validate the goodness of a certain solution is the Estimated Distance to the (global) Minimum (EDM): an EDM value closer to zero indicates a successful fitting function minimization. These features motivate our election of the Minuit curve fitting program to perform the RCs analysis of UGC 6446 and UGC 7524. 

\begin{figure*}
\centering
\includegraphics[width=1.0\hsize]{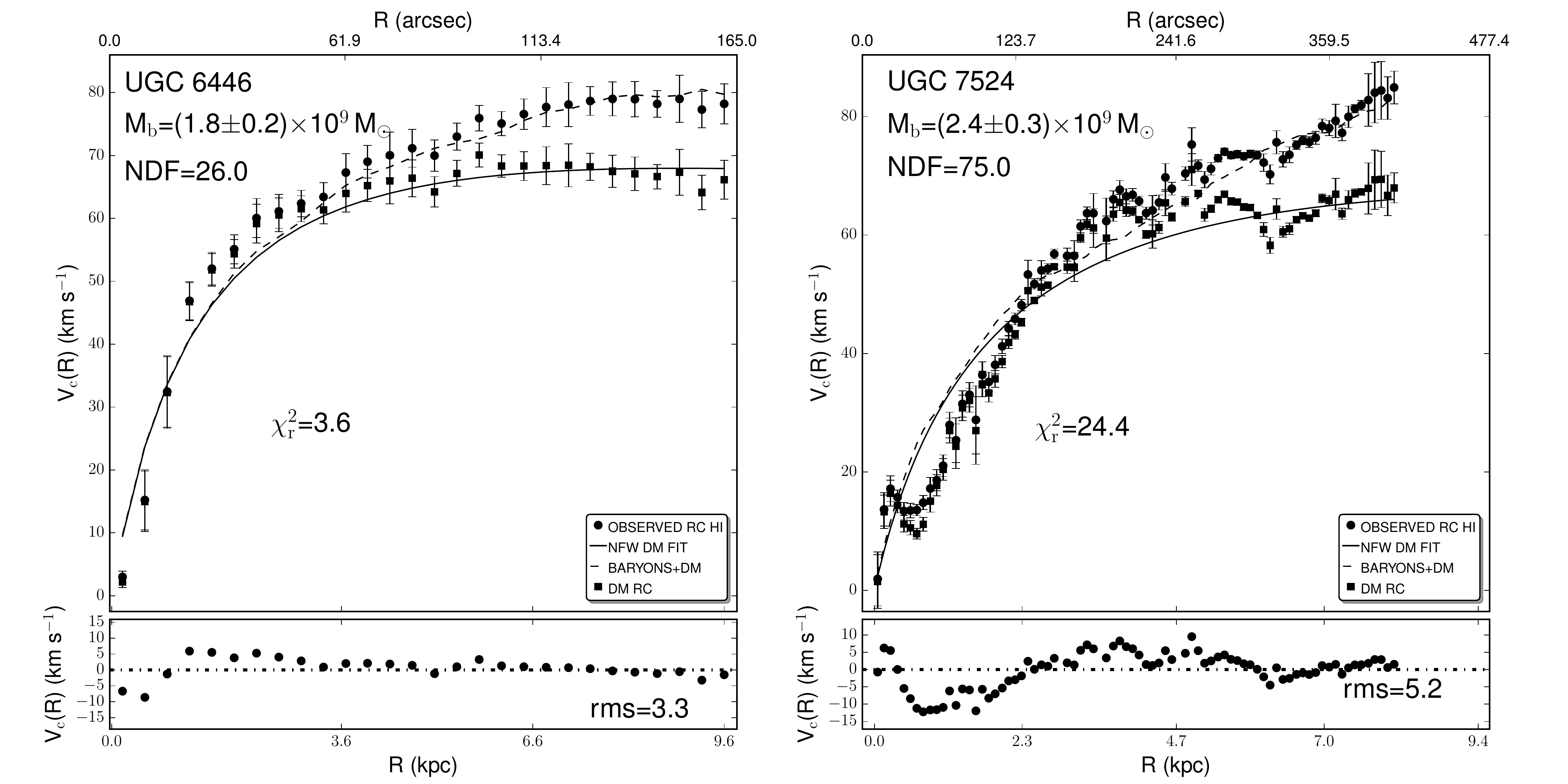}
\caption[f7.pdf]{NFW DM halo fitting results for UGC 6446 and UGC 7524. 
In the top panels are indicated the number of degrees of freedom (NDF), 
the reduced $\chi_{\rm r}^2$ normalized by the respective NDF and the baryonic
masses of both galaxies as estimated in section~\ref{sec:sc3}.
The solid curve is the NFW fit to the DM RCs and the dashed curve is the square
root of the sum of the squares of the fitted DM RCs and the 
stellar and HI+He+metals RCs (i.e., the total circular velocity). 
The residuals in the bottom panels are the difference between the observed HI RCs
and the total circular velocities (dashed curves).
~\label{fig7}}
\end{figure*}

\begin{figure*}
\centering
\includegraphics[width=1.0\hsize]{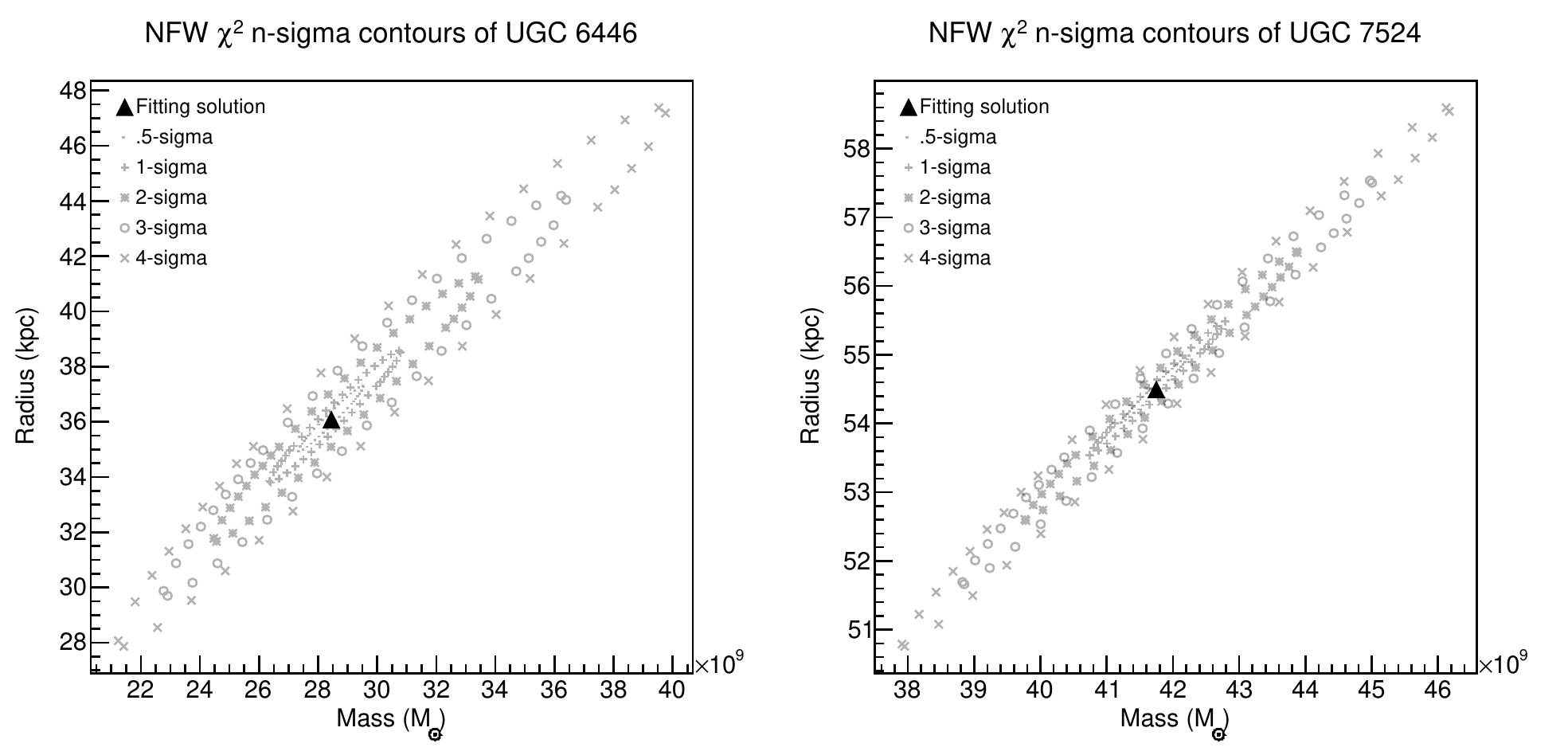}
\caption[f8.pdf]{NFW DM halo $\chi^2$ n-sigma contours for UGC 6446 and UGC 7524. 
The 0.5, 1.0, 2.0, 3.0 and 4.0 sigma contours are displayed and the fitting 
solution is showed by means of a black filled triangle. The scanned ranges for 
DM halo mass and radius for UGC 6446 and UGC 7524 are exhibited and their values
for global minimum are listed in Table~\ref{tab:tb3}.
~\label{fig8}}
\end{figure*}

\begin{figure*}
\centering
\includegraphics[width=1.0\hsize]{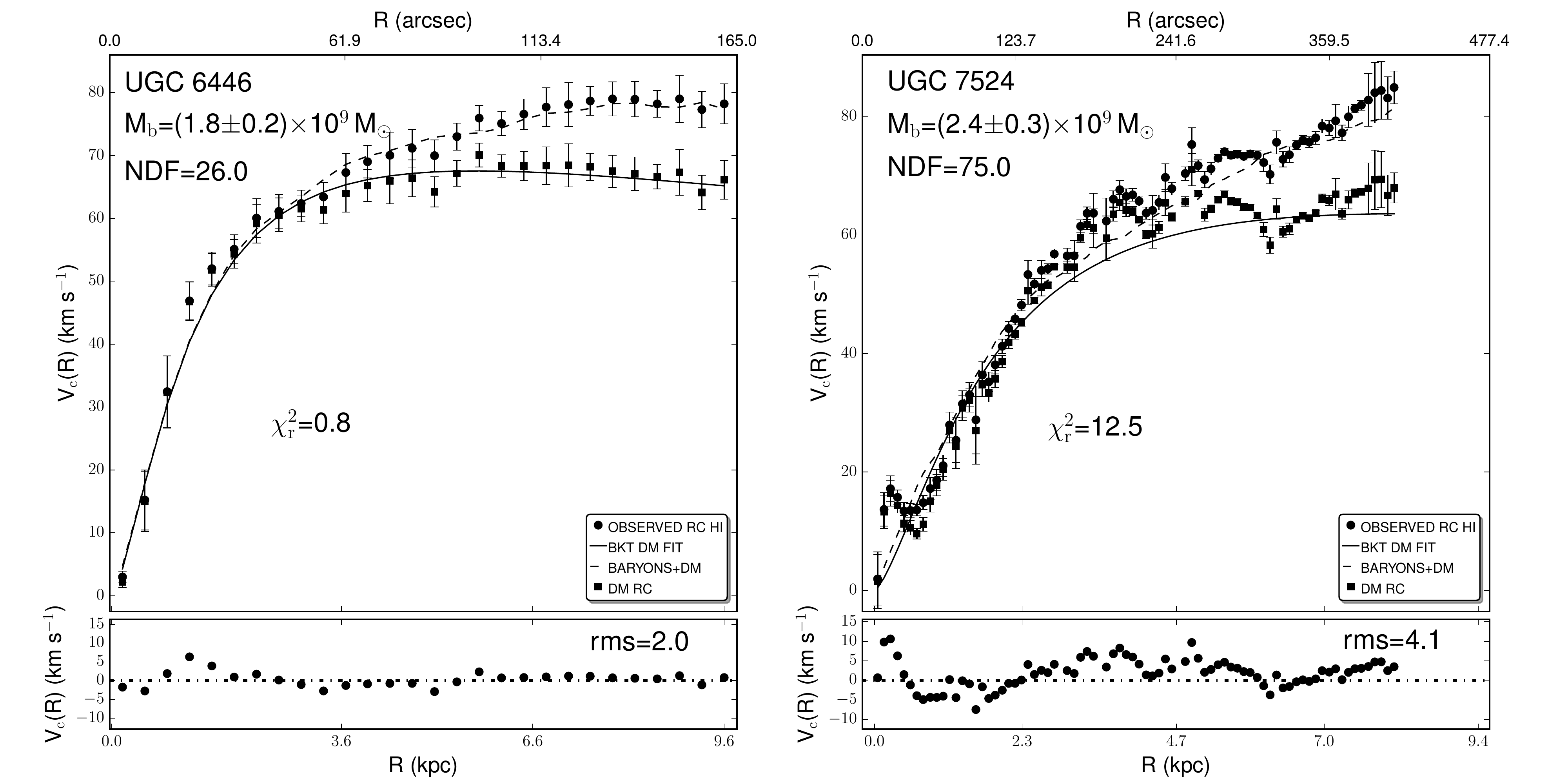}
\caption[f9.pdf]{BKT DM halo fitting results for UGC 6446 and UGC 7524.
The description of the panels is the same as in Figure~\ref{fig7}.
~\label{fig9}}
\end{figure*}

\begin{figure*}
\centering
\includegraphics[width=1.0\hsize]{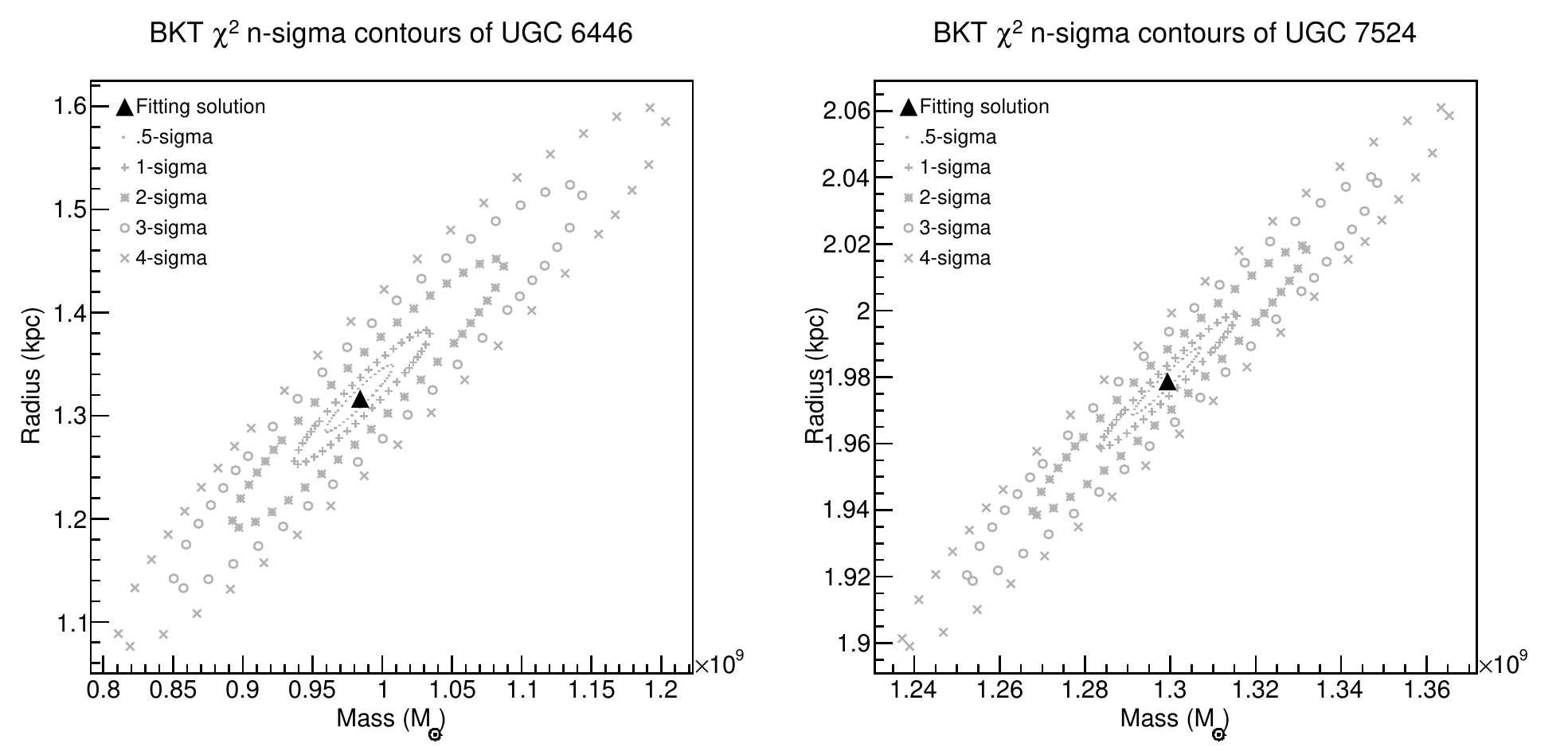}
\caption[f10.pdf]{BKT DM halo $\chi^2$ n-sigma contours for UGC 6446 and UGC 7524.
The description of the panels is the same as in Figure~\ref{fig8}.
~\label{fig10}}
\end{figure*}

\begin{figure*}
\centering
\includegraphics[width=1.0\hsize]{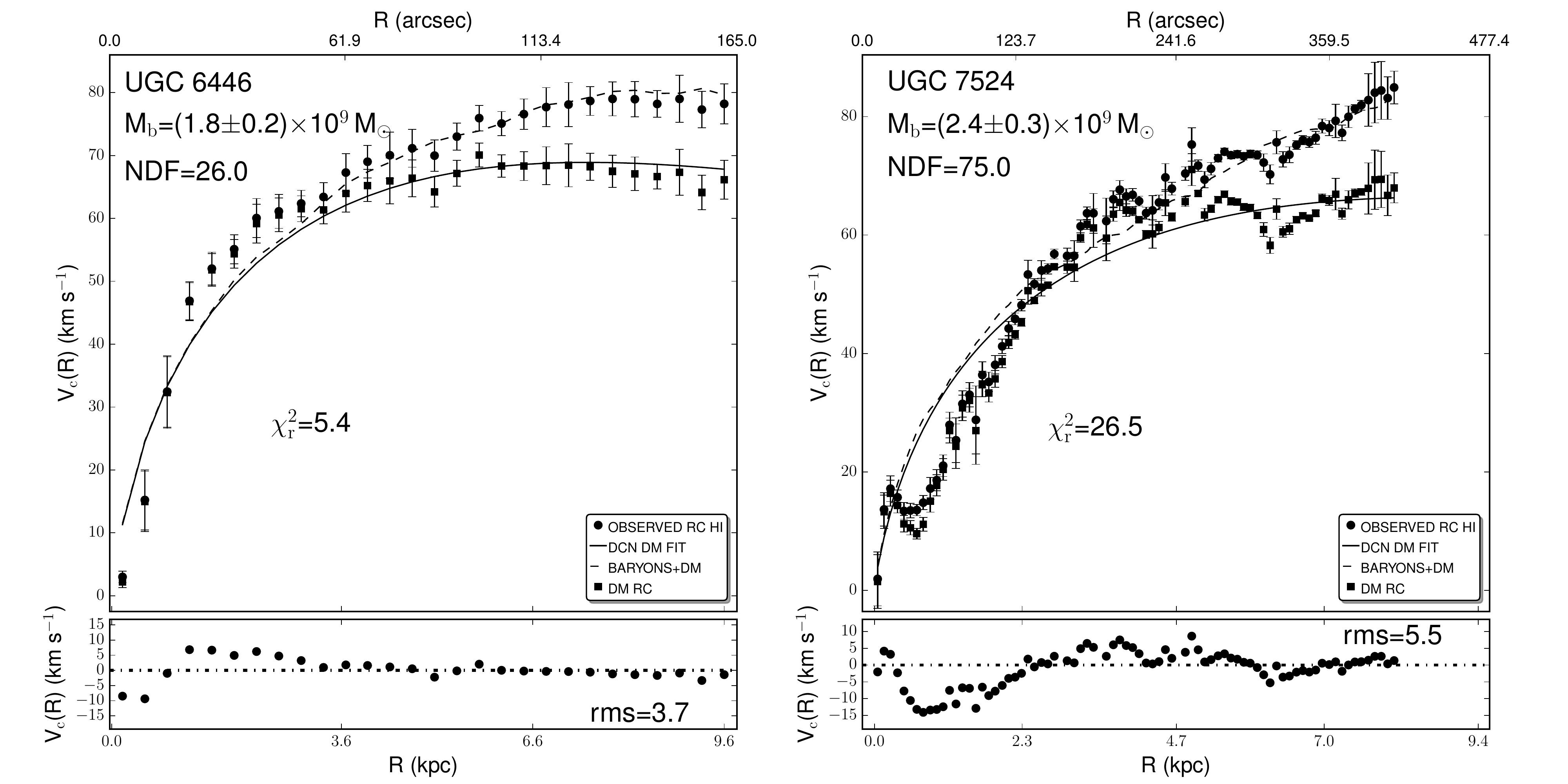}
\caption[f11.pdf]{DCN DM halo fitting results for UGC 6446 and UGC 7524.
The description of the panels is the same as in Figure~\ref{fig7}.
~\label{fig11}}
\end{figure*}

\begin{figure*}
\centering
\includegraphics[width=1.0\hsize]{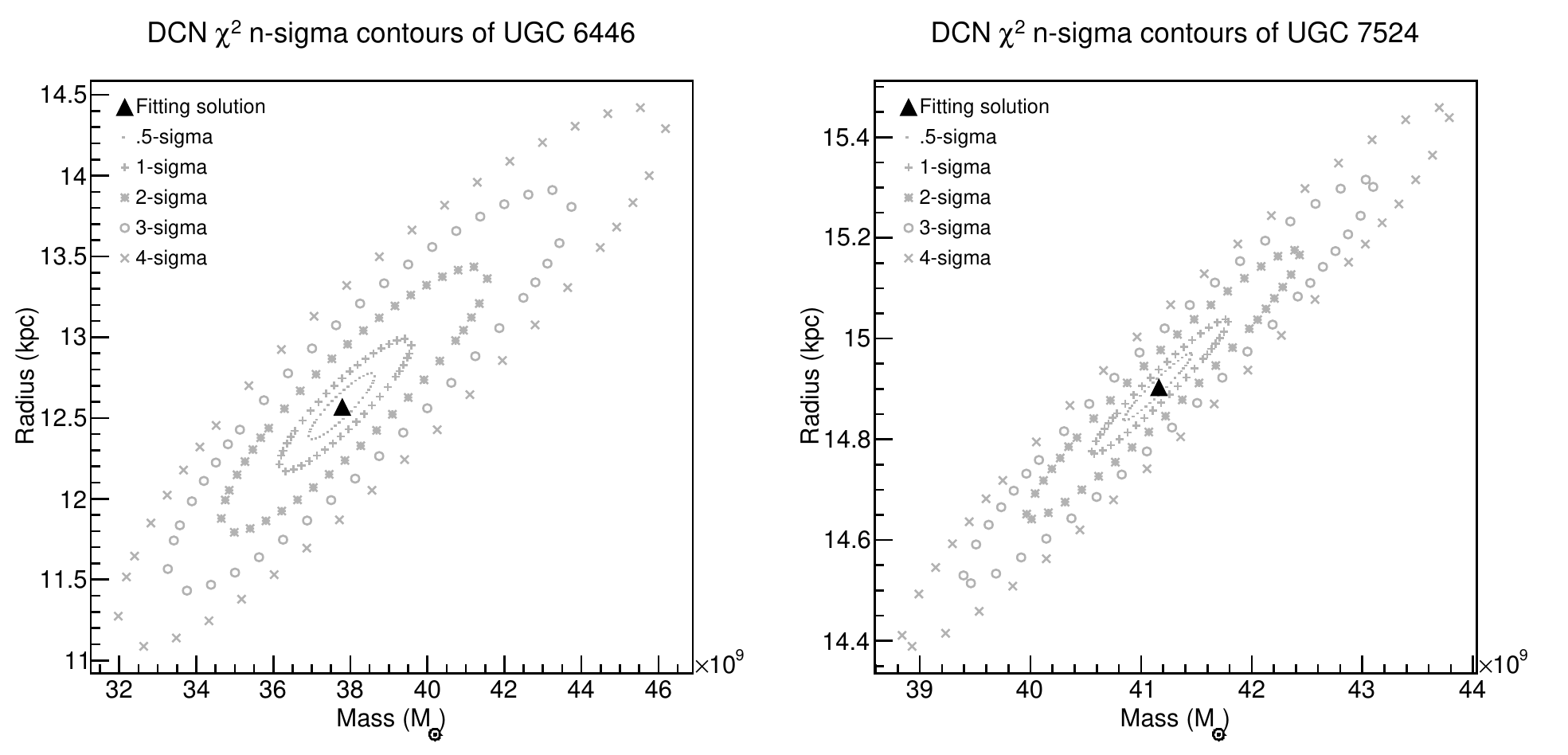}
\caption[f12.pdf]{DCN DM halo $\chi^2$ n-sigma contours for UGC 6446 and UGC 7524.
The description of the panels is the same as in Figure~\ref{fig8}.
~\label{fig12}}
\end{figure*}

\begin{figure*}
\centering
\includegraphics[width=1.0\hsize]{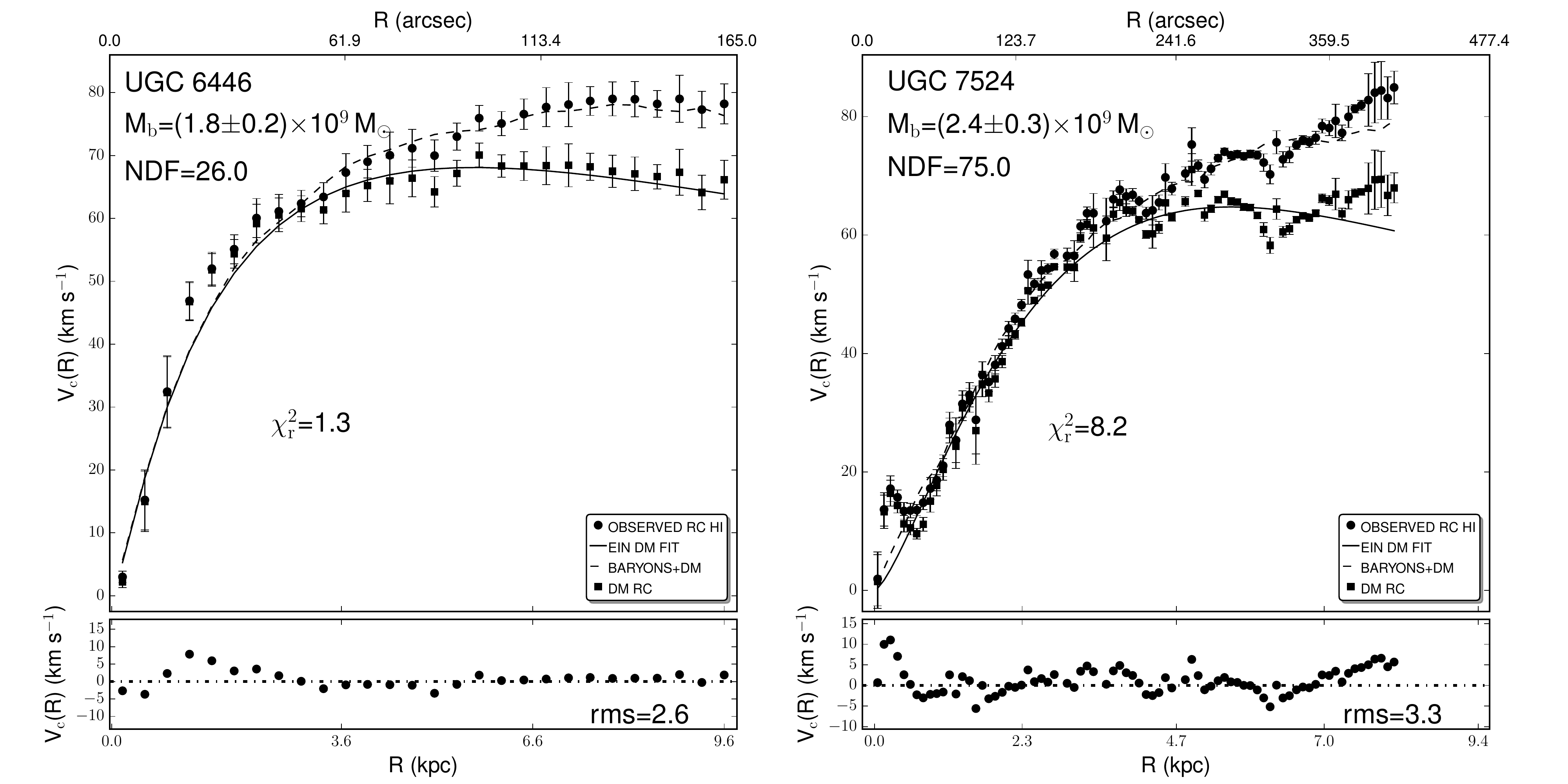}
\caption[f13.pdf]{EIN DM halo fitting results for UGC 6446 and UGC 7524.
The description of the panels is the same as in Figure~\ref{fig7}.
~\label{fig13}}
\end{figure*}

\begin{figure*}
\centering
\includegraphics[width=1.0\hsize]{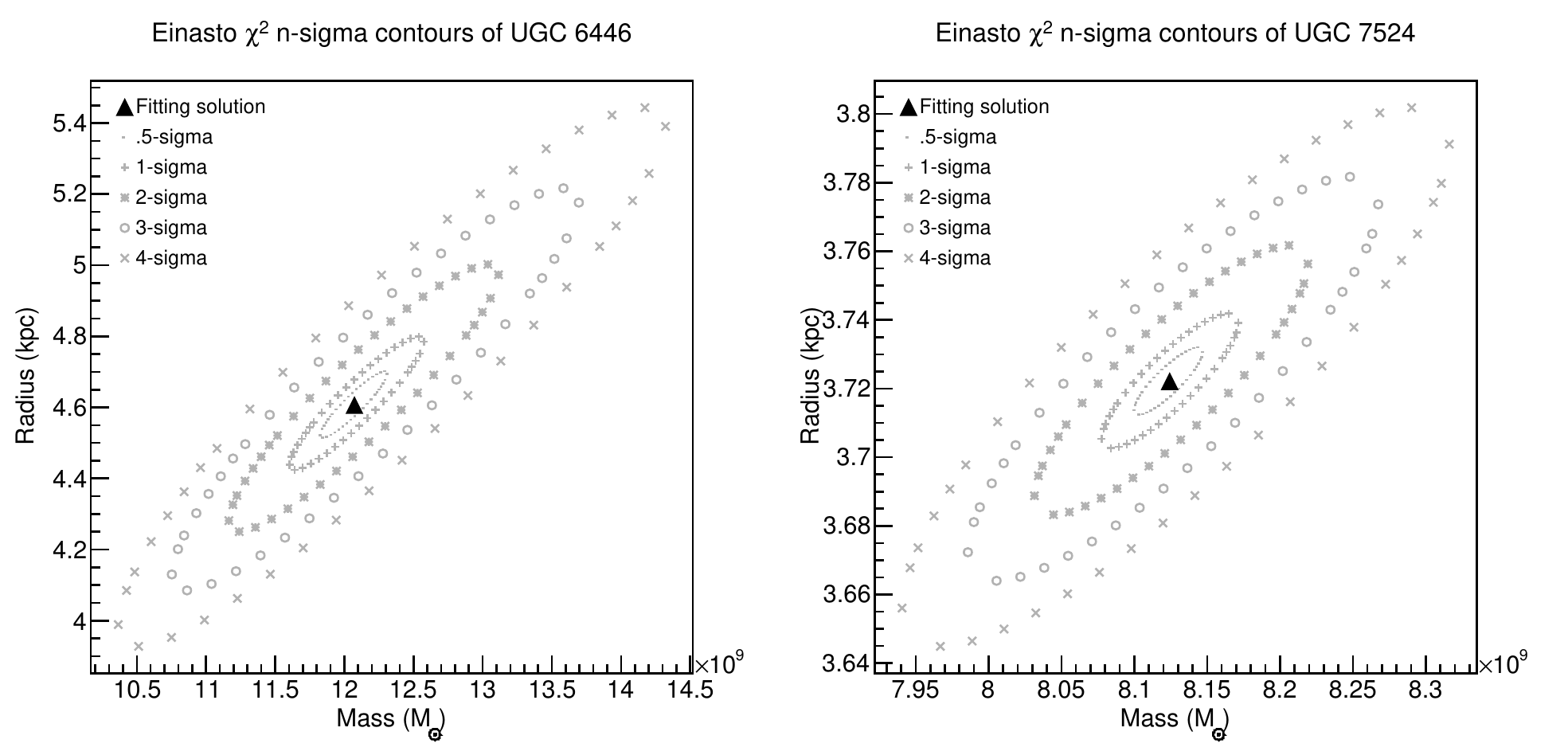}
\caption[f14.pdf]{EIN DM halo $\chi^2$ n-sigma contours for UGC 6446 and UGC 7524.
The description of the panels is the same as in Figure~\ref{fig8}.
~\label{fig14}}
\end{figure*}

\begin{figure*}
\centering
\includegraphics[width=1.0\hsize]{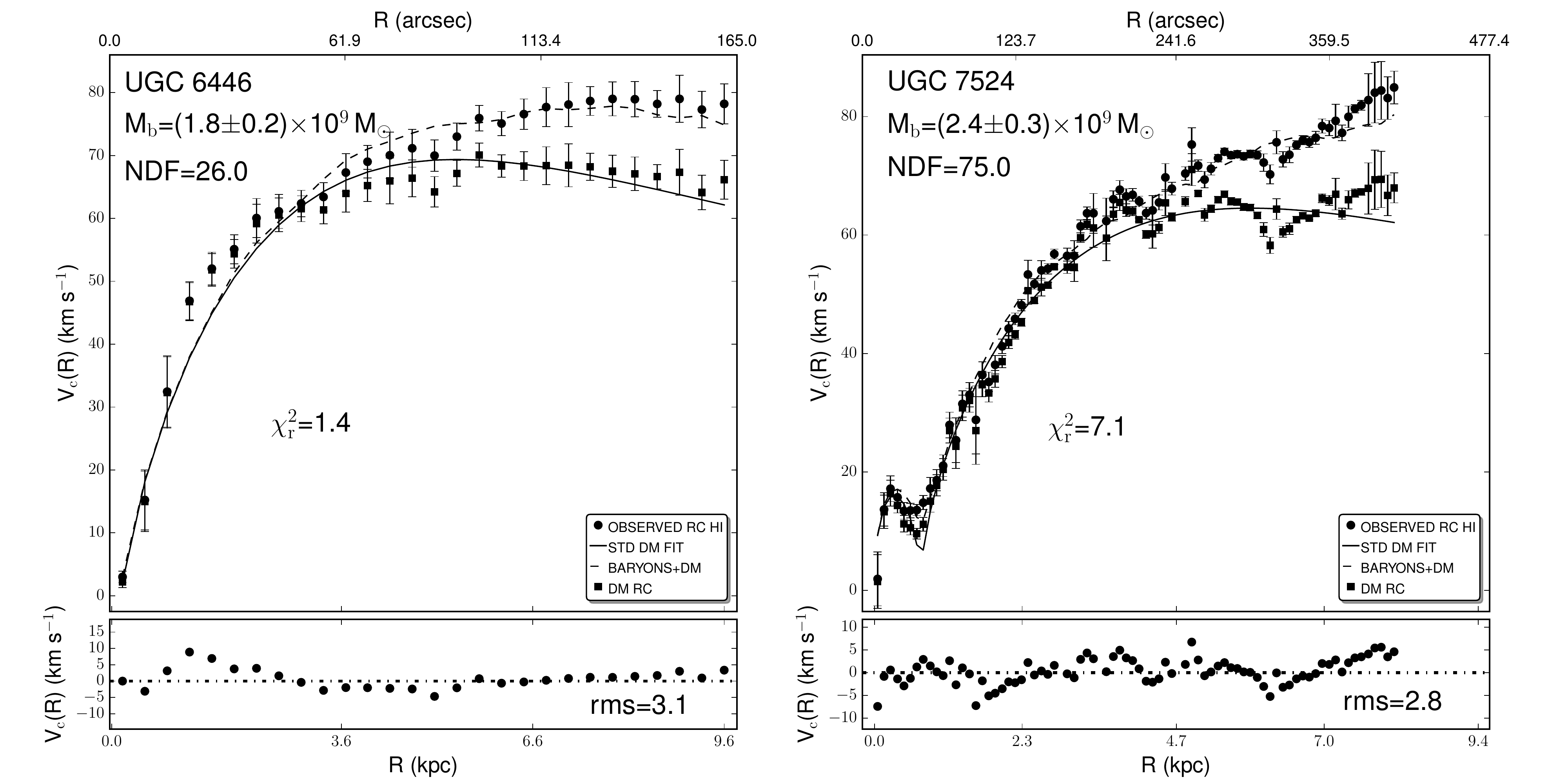}
\caption[f15.pdf]{STD DM halo fitting results for UGC 6446 and UGC 7524.
The description of the panels is the same as in Figure~\ref{fig7}.
~\label{fig15}}
\end{figure*}

\begin{figure*}
\centering
\includegraphics[width=1.0\hsize]{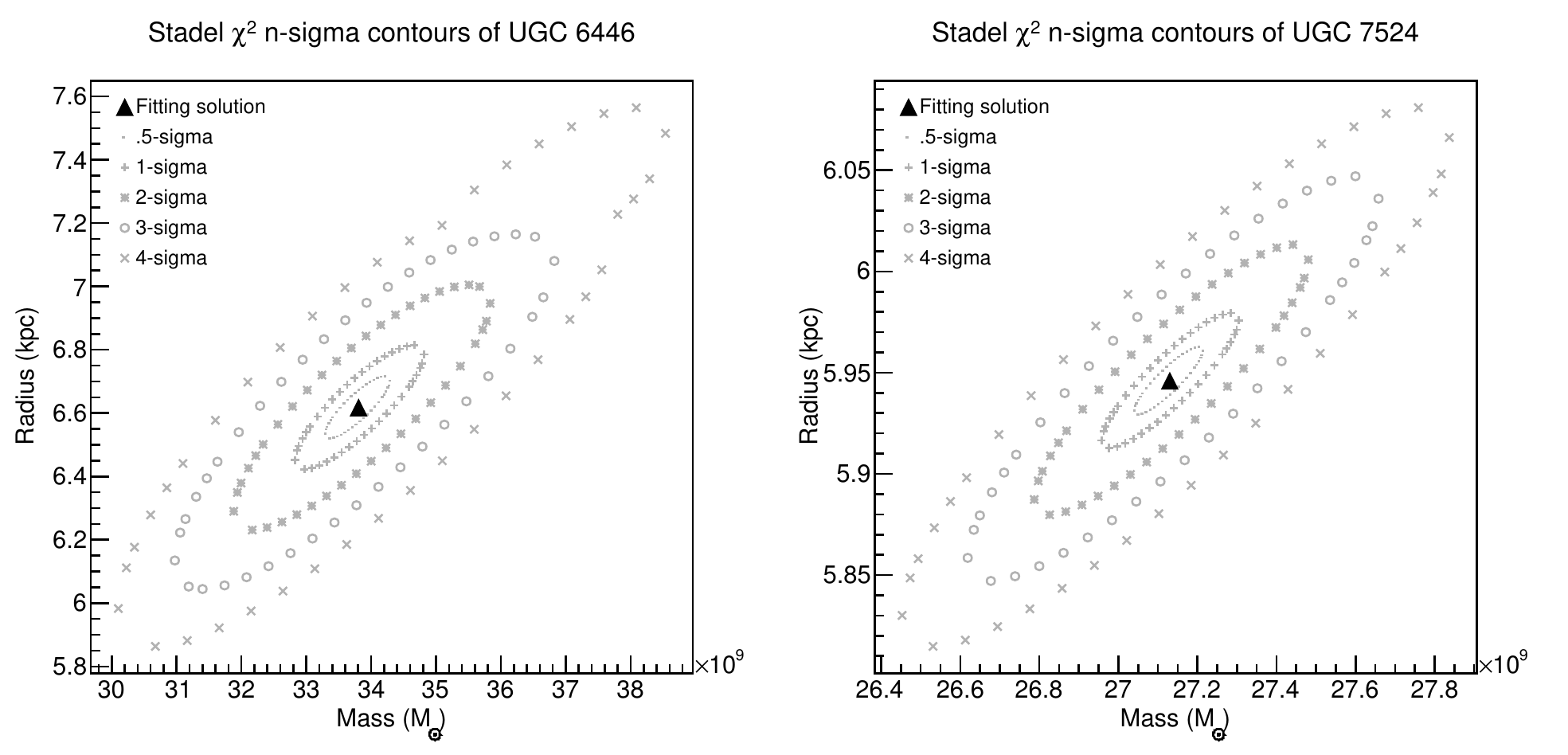}
\caption[f16.pdf]{STD DM halo $\chi^2$ n-sigma contours for UGC 6446 and UGC 7524.
The description of the panels is the same as in Figure~\ref{fig8}.
~\label{fig16}}
\end{figure*}

The DM models used in the present study are the same employed in RP15: the 
\citet{Navarro1996} DM halo (NFW), the \citet{Burkert1995} DM halo (BKT), 
the \citet{DiCintio2014} DM halo (DCN), the \citet{Einasto1965} DM halo 
(EIN) and the \citet{Stadel2009} DM halo (STD). We use the \citet{Dutton2014}
prescription for the concentration parameter and the approximation 
of \citet{MacArthur2003} where the recipe of \citet{Ciotti1999} is not valid
for the EIN DM halo. The NFW, BKT and DCN DM halo profiles have two free parameters,
namely, the halo mass and radius, whereas the EIN and STD DM halo profiles have an 
additional parameter characterizing the shape of the DM halos profile.  As in RP15 
we fix the third parameter of EIN and STD allowing solely the variation of the halo
mass and radius. The EIN shape parameter was set to 1.8 and 
0.72 respectively for UGC 6446 and UGC 7524, whereas the STD 
shape parameter was fixed to 3.7 and 3.1 
separately for UGC 6446 and UGC 7524. These values are the product of an extensive 
test search that fixes the third parameter to a new value at each run of the two 
free parameters. We fix the third parameter whenever the reduced 
$\chi^2$, EDM and $\chi^2$ n-sigma contours indicate that the RCs fitting process
reached the expected global minimum. The range of admitted values of the DM halos 
masses and radii are [10$^8$, 10$^{13}$] M$_{\odot}$ and [0.1, 100] 
kpc, respectively. The main results are displayed in 
figures~\ref{fig7} -- \ref{fig16} and are listed in Table~\ref{tab:tb3}. 
We provide in pictorial form the solutions of NFW, BKT, DCN, EIN and STD DM models 
together with the corresponding $\chi^2$ n-sigma contours and the range of variation
of the free parameters. 

\begin{table}
\centering
\begin{threeparttable}
\caption{DM RCs fit results for UGC 6446 and UGC 7524.}
\label{tab:tb3}
\begin{tabular}{p{1.0cm}p{1.5cm}p{1.2cm}p{0.9cm}p{0.4cm}p{0.9cm}}
\hline
\hline
DMMs\tnote{(11)} & M$_{\rm tot}$ (M$_{\odot}$)\tnote{(12)} & M$_{\rm h}$ 
(M$_{\odot}$)\tnote{(12)} & R$_{\rm h}$\tnote{(13)} & $\chi_r^2$\tnote{(14)} & EDM\tnote{(15)}\\
\hline
\hline
NFW & 3.0$\pm$0.2 & 2.8$\pm$0.2 & 36.0$\pm$2.4 & 3.6 & 2.5$\,\times\,10^{-7}$\\
\hline
BKT & 0.3$\pm$0.03 & 0.1$\pm$0.005 & 1.3$\pm$0.07 & 0.8 & 2.6$\,\times\,10^{-7}$\\
\hline
DCN & 4.0$\pm$0.2 & 3.8$\pm$0.2 & 13.0$\pm$0.4 & 5.4 & 7.9$\,\times\,10^{-9}$\\
\hline
EIN & 1.4$\pm$0.07 & 1.2$\pm$0.05 & 4.6$\pm$0.2 & 1.3 & 2.4$\,\times\,10^{-8}$\\
\hline
STD & 3.6$\pm$0.1 & 3.4$\pm$0.1 & 6.6$\pm$0.2 & 1.4 & 4.1$\,\times\,10^{-13}$\\
\hline
\hline
\hline
\hline
NFW & 4.4$\pm$0.1 & 4.2$\pm$0.1 & 55.0$\pm$1.0 & 24.4 & 4.3$\,\times\,10^{-9}$\\
\hline
BKT & 0.4$\pm$0.02 & 0.1$\pm$0.002 & 2.0$\pm$0.02 & 12.5 & 2.5$\,\times\,10^{-11}$\\
\hline
DCN & 4.4$\pm$0.1 & 4.1$\pm$0.06 & 15.0$\pm$0.1 & 26.5 & 3.8$\,\times\,10^{-9}$\\
\hline
EIN & 1.1$\pm$0.03 & 0.8$\pm$0.005 & 3.7$\pm$0.02 & 8.2 & 5.0$\,\times\,10^{-8}$\\
\hline
STD & 3.0$\pm$0.04 & 2.7$\pm$0.02 & 6.0$\pm$0.03 & 7.1 & 1.8$\,\times\,10^{-12}$\\
\hline
\hline
\end{tabular}
\begin{tablenotes}
\item[(11)] DM halo models.
\item[(12)] Total masses and DM halo masses normalized by $10^{10}$.
\item[(13)] DM halo radii in kpc.
\item[(14)] Reduced $\chi^2$ normalized by the number of degrees of freedom.
\item[(15)] Estimated distance to the minimum (see text).
\end{tablenotes}
\end{threeparttable}
\end{table}

From the reduced $\chi^2$ values, EDM, the $\chi^2$ n-sigma contours, 
the parameters scan intervals, and the fits visualisations it is evident that the NFW 
DM model adjusts the DM RCs of UGC 6446 and UGC 7524 better than the DCN DM halo, 
nevertheless EIN, BKT and STD DM models reproduce more adequately the DM RCs of UGC 6446 
and UGC 7524. The BKT DM halo performs better than any other DM halo 
used in this analysis for UGC 6446, however for UGC 7524 the STD solution is the best. The DCN DM model 
does not reproduce the DM RCs of UGC 6446 and UGC 7524 properly. The inner slope of the DCN DM halo is 
close to a cuspy profile with values near to 0.38 and 0.36 for UGC 6446 and UGC 7524,
respectively. In general the two exponential DM models used in this work, 
EIN and STD behave better for UGC 7524 or nearly at the same level for UGC 6446 than the BKT DM model. 
This trend confirms some of the conclusions attained in RP15.

\begin{table}
\centering
\begin{threeparttable}
\caption{Comparison of some of our results with the findings of SW11 and VB01.}
\label{tab:tb5}
\begin{tabular}{p{1.1cm}p{1.8cm}p{0.4cm}p{2.0cm}p{1.1cm}}
\hline
\hline
Authors & V$_{200}$ (km s$^{-1}$) & c$_{200}$ & $\rho_{\rm h}$ (M$\odot$ pc$^{-3}$) & r$_{\rm h}$ (kpc)\\
\hline
\hline
PW\tnote{(16)} & 58.2 & 12.0 & (100.0$\pm$4.0)$\times$10$^{-3}$ & 1.3$\pm$0.07\\
\hline
PW & 57.4 & 11.5 & (34.0$\pm$2.0)$\times$10$^{-3}$ & 2.0$\pm$0.02\\
\hline
SW11\tnote{(17)} & ------ & ------ & (120.0$\pm$20.0)$\times$10$^{-3}$ & 0.95$\pm$0.1\\
\hline
SW11 & ------ & ------ & (59.0$\pm$7.0)$\times$10$^{-3}$ & 1.5$\pm$0.1\\
\hline
VB01\tnote{(18)} & 52.0 & 17.4 & ------ & ------\\
\hline
VB01 & 72.0 & 8.5 & ------ & ------\\
\hline
\hline
\end{tabular}
\begin{tablenotes}
\item[(16)] Present work, the first row is for UGC 6446 and the second for UGC 7524 for each authors listed in the table.
\item[(17)] Pseudo-isothermal DM model.
\item[(18)] NFW DM halo.
\end{tablenotes}
\end{threeparttable}
\end{table}

We compare the results of the BKT DM halo and the NFW DM halo for UGC 6446 and UGC 7524 
with the findings of \citet{vandenBosch2001} (VB01) and \citet{Swaters2011} (SW11). 
VB01 studied 20 dwarf galaxies, including the two galaxies object of the present study, 
to derive the concentration parameters and the circular velocities where the contrast 
between the universe mean density and the DM halo average density is 200 (i.e. V$_{200}$).
The authors used the NFW DM model and derived the stellar and gas discs from R-band surface
brightness profiles and HI profiles respectively, assuming a constant M/L. SW11 analysed the
HI RCs of nearly the same group of galaxies of VB01 to determine the DM central densities and
scale radii of the entire sample. The authors employed the Pseudo-Isothermal DM model and 
obtained the baryonic discs as in VB01. The results of SW11, VB01 and the outcomes of the 
present study for UGC 6446 and UGC 7524 are listed in Table~\ref{tab:tb5}. 
From the values of V$_{200}$ and c$_{200}$ reported in Table~\ref{tab:tb5} 
it is apparent the concordance with VB01, even though the adopted definition
of the NFW DM halo concentration in our work differs from that used in VB01. The central 
densities and core radii of the BKT DM model in the present study are in general concordant 
with the corresponding quantities of the Pseudo-isothermal DM halo employed in SW11, even 
though a direct comparison is difficult because of the distinct DM model used, the different
assumptions on the M/L ratio and the diverse methods employed in our analysis to obtain the 
baryonic content.

\section{Possible problems in DM content estimates}\label{sec:sc5}

The current section addresses the principal issues related to the determination of the DM 
and total mass of UGC 6446 and UGC 7524 such as for instance the proper estimation of 
possible non circular motions, the spectral types included in the stellar RCs of UGC 6446
and UGC 7524 and the gaseous phases contained in the HI+He+metals RCs of 
both galaxies. These three points are of fundamental importance to appraise the reliability 
of the total observed HI RCs and the stellar and neutral gas plus helium 
plus metals RCs of UGC 6446 and UGC 7524. In the next paragraphs we discuss separately the
three topics outlined above and the respective implications.

\begin{enumerate}

\item In section~\ref{sec:sc3} we estimate that the degree of kinematic lopsidedness of both galaxies 
is negligible consistently with the findings of \citet{vanEymeren2011a}. The morphological lopsidedness of UGC 6446
and UGC 7524 as derived by \citet{vanEymeren2011b} through harmonic analysis of the total HI density map of both 
galaxies is comparatively higher than the kinematic lopsidedness estimated in the same paper (see their table A.1). 
The authors conclude 
that for the sample of analysed galaxies the most probable cause of the observed lopsidedness is interaction with 
nearby companions. In the particular case of UGC 6446 and UGC 7524, the hypothesis of tidal encounters as the 
generating mechanism of lopsidedness seems reasonable since both galaxies are members of two different agglomeration
of galaxies. UGC 6446 belongs to the Ursa Major group of galaxies a conglomerate of 79 recognized members at a distance
of $\sim$19 Mpc and with a velocity dispersion of 150 km s$^{-1}$ \citep{Verheijen2001}. UGC 7524 is part of the Coma I cloud, 
an aggregation of 132 galaxies at a distance less than $\sim$5 Mpc and with a velocity dispersion of 
307 km s$^{-1}$ \citep{Boselli2009}. The larger velocity dispersion of the Coma I cloud indicates a higher rate of interactions between 
the galaxies in this group with respect to the galaxies in the Ursa Major agglomerate. As a consequence, we expect that
objects in groups with a high velocity dispersion manifest a considerable amount of lopsidedness 
when compared to agglomerations of galaxies with small velocity dispersion. This conclusion is partially
confirmed by the results of \citet{vanEymeren2011b} because the morphological lopsidedness of UGC 6446 is smaller than that of UGC 7524. 
\citet{vanEymeren2011a} and our estimates of the kinematic lopsidedness of both galaxies clearly contradict the expected
trend reported above as one can see from Table~\ref{tab:tb2}. \citet{Angiras2007} perform harmonic analysis 
of the HI total density maps and velocity fields of UGC 6446 and other 10 galaxies of the Ursa Major group. The authors find a 
kinematic lopsidedness for UGC 6446 of nearly a factor of ten higher with respect to \citet{vanEymeren2011a} and our estimates. We 
are not aware of a work similar to that of \citet{Angiras2007} for UGC 7524, however we might guess for this galaxy a difference 
comparable or higher to that estimated for UGC 6446 between the values of kinematic lopsidedness computed by \citet{vanEymeren2011a} 
and this article and an hypothetical study employing the methods of \citet{Angiras2007}. The 
reported discrepancy is not surprising and relies on the different methods used by the quoted authors to compute the 
kinematic lopsidedness. The procedure of \citet{Jog2002}, applied by \citet{vanEymeren2011a} and this study, measures the intensity 
of the lopsided perturbation as the difference between the maximum velocities of the receding and approaching 
side of the RC normalized by twice the maximum of the circular velocity of both sides of the same RC. This technique is a
very powerful tool to determine the degree of kinematic lopsidedness within the angular sector where the RC is derived. On
the other hand, the methodology employed by \citet{Angiras2007} searches for possible deviations from circular 
(or projected elliptical) motions across the entire 2D velocity field and interprets these deviations as Fourier modes. This latter 
technique quantifies the kinematic lopsidedness of the whole radial velocity map, not only of the region near the kinematic major 
axis where the RC is determined. The disparity in the measured values of the kinematic lopsidedness obtained by the application of 
the two different approaches are well accounted by the papers quoted above at least in the case of UGC 6446.
In this work we are interested in the local effect of the kinematic lopsidedness on the observed RCs of UGC 6446 and 
UGC 7524, consequently the application of the method of \citet{Jog2002} seems adequate to 
our purpose and as a consequence the kinematic lopsidedness has a negligible effect in the derivation of the observed RCs of the two 
galaxies studied in this paper as established in section~\ref{sec:sc3}.

\item The stellar discs of UGC 6446 and UGC 7524 harbour the variety of spectral types (together with gas, dust 
and interstellar medium) capable to produce the observed photometry available in the literature of both galaxies covering an approximate
wavelength range from the far ultraviolet ($\sim$0.1 $\mu$m) to the far infrared ($\sim$10$^3$ $\mu$m). The SPS approach we employed in 
this work (e.g. \citet{bru03}) to obtain the stellar mass maps, surface densities and RCs of UGC 6446 and UGC 7524 encompasses the necessary
values of temperatures, masses, ages and also some crucial phases of stellar evolution such as core convective overshooting, prescriptions 
for TP-AGB stars and mass-loss. As a consequence SPS models properly reproduce the stellar populations that actually inhabit
the stellar disc of both galaxies. On the other hand, stellar rotation and binary evolution are not included in the SPS model we used to build the surface densities and these omissions could in principle affect the determination of the total mass. As we emphasised in the introduction stellar rotation produces bluer SPS colors and higher luminosities with respect to SPS without rotation. The net result of using SPS without rotation could lead to an overestimation of the stellar population masses by a factor of two \citep{Vazquez2007}. The authors caution that their work is preparatory and much more efforts 
are needed to evaluate the complete reliability of their main findings. As established by \citet{Eldrige2008} the effect on SPS from binary evolution of non rotating companions are comparable to some degree to the consequences of rotating single 
stars. Nevertheless, in the present investigation we do not incorporate UV observations of early-type galaxies, and 
therefore the effects of binary evolution should be negligible \citep{Hernandez2013}. Summarising the estimated stellar 
RCs of UGC 6446 and UGC 7524 should reproduce the actual stellar RCs with reasonable confidence.

\item The total mass of atomic hydrogen of UGC 6446 and UGC 7524 according to \citet{Swaters2002} is 
$\sim$1.3$\times$10$^9$ M$_{\odot}$ and $\sim$1.0$\times$10$^9$ M$_{\odot}$ inside a radius of $\sim$11 kpc and $\sim$9 kpc
(adopting a distance of 12 and 3.5 Mpc) respectively. \citet{Leroy2005} accomplished a CO survey of the central regions of 
121 dwarf galaxies to measure the amount of H$_2$ in these objects and reported no detection of CO emission for UGC 6446. 
This result is confirmed by the study of \citet{Watson2012} that analyse the molecular gas surface density of 20 galaxies 
of morphological types Sd-Sdm. The mass of molecular hydrogen in the central region of UGC 7524 
is 2.0$\times$10$^5$ M$_{\odot}$ according to the estimation of \citet{Boker2011}. The global mass of atomic and molecular hydrogen of 
UGC 7524 as estimated by \citet{Stark1987} is 1.2$\times$10$^9$ M$_{\odot}$ and 2.0$\times$10$^8$ M$_{\odot}$
separately. From these data we conclude that the neutral gas plus helium plus metals RCs of UGC 6446 and 
UGC 7524 derived in this work represent a good approximation to the total gas mass of these two galaxies given that other 
gaseous phases constitute a negligible mass amount to the total gas mass budget. In addition our neutral gas mass estimates 
of (1.2$\pm$0.2)$\times$10$^9$ M$_{\odot}$ and (1.3$\pm$0.3)$\times$10$^9$ M$_{\odot}$, for UGC 6446 and 
UGC 7524 respectively, are concordant with those of \citet{Swaters2002}.

\end{enumerate}

\section{Conclusions}\label{sec:sc6}

In this article we estimate the total mass of UGC 6446 and UGC 7524 analysing high resolution HI RCs
derived from full resolution HI velocity fields. The latter were observed with the Westerbork Synthesis Radio Telescope
under the WHISP project. The chief findings about the total mass content of UGC 6446 and UGC 7524 are itemised below.

\begin{itemize}
\item The sum of the stellar and neutral gas plus helium plus metals RCs of UGC 6446 and UGC 7524 is 
unable to reproduce the total observed
HI RCs of both galaxies. This fact suggests the necessity of substantial amounts of DM.\\
\item Cored DM models adjust more satisfactory the DM RCs of the galaxies examined in this work with respect to cuspy
DM halos.\\
\item The cored DM halo BKT produces the best fit of the UGC 6446 DM RC, whereas the STD cored DM halo
provides the best performance for the DM RC of UGC 7524.\\
\end{itemize}

In general the two exponential DM models, EIN and STD, with only two free parameters
give results similar or better than the cored BKT DM halo. This tendency partially supports the results of RP15 and encourages
the study of these two DM models in future works to increase the statistical robustness of our conclusions.

\section*{Acknowledgements}

We acknowledge the anonymous referee for constructive comments and suggestions.
M. R. thanks the DGAPA-PAPIIT IN103116 from UNAM and CY-253085 from CONACYT 
projects for financial support during the production of this article. We have
made use of the WSRT on the Web Archive. The Westerbork Synthesis Radio 
Telescope is operated by the Netherlands Institute for Radio Astronomy ASTRON, 
with support of NWO. We acknowledge the usage of the HyperLeda database 
(http://leda.univ-lyon1.fr). This research has made use of the NASA/IPAC 
Extragalactic Database (NED) which is operated by the Jet Propulsion Laboratory, 
California Institute of Technology, under contract with the National Aeronautics 
and Space Administration. This research has made use of the SIMBAD database,
operated at CDS, Strasbourg, France. Based on observations made with the NASA/ESA 
Hubble Space Telescope, and obtained from the Hubble Legacy Archive, which is a 
collaboration between the Space Telescope Science Institute (STScI/NASA), the Space 
Telescope European Coordinating Facility (ST-ECF/ESA) and the Canadian Astronomy Data
Centre (CADC/NRC/CSA). This research has made use of the NASA/ IPAC Infrared Science 
Archive, which is operated by the Jet Propulsion Laboratory, California Institute 
of Technology, under contract with the National Aeronautics and Space Administration.
Funding for SDSS-III has been provided by the Alfred P. Sloan Foundation, 
the Participating Institutions, the National Science Foundation, and the U.S. Department 
of Energy Office of Science. The SDSS-III web site is http://www.sdss3.org/.

SDSS-III is managed by the Astrophysical Research Consortium for the Participating Institutions
of the SDSS-III Collaboration including the University of Arizona, the Brazilian Participation 
Group, Brookhaven National Laboratory, Carnegie Mellon University, University of Florida, the 
French Participation Group, the German Participation Group, Harvard University, the Instituto 
de Astrofisica de Canarias, the Michigan State/Notre Dame/JINA Participation Group, Johns Hopkins
University, Lawrence Berkeley National Laboratory, Max Planck Institute for Astrophysics, Max 
Planck Institute for Extraterrestrial Physics, New Mexico State University, New York University,
Ohio State University, Pennsylvania State University, University of Portsmouth, Princeton 
University, the Spanish Participation Group, University of Tokyo, University of Utah, Vanderbilt
University, University of Virginia, University of Washington, and Yale University.










\bsp	
\label{lastpage}
\end{document}